\documentclass[usenatbib]{aa}  

\usepackage{graphicx}
\usepackage{graphicx}	% Including figure files
\usepackage{amsmath}	% Advanced maths commands
\usepackage{amssymb}	% Extra maths symbols
\usepackage{booktabs}
\usepackage{epsfig}
\usepackage{tablefootnote}
\usepackage{epstopdf}
\usepackage{xcolor}
\usepackage[colorlinks=true,citecolor=magenta,linkcolor=blue]{hyperref}
\usepackage{txfonts}
\begin{document}

   \title{Multi-wavelength Analysis and Modeling of OJ 287 During 2017-2020}
 \titlerunning{ Temporal and Spectral study of OJ 287}
   \authorrunning{Prince et al.}

%   \author{G. Wuchterl
%          \inst{1}
%          \and
%          C. Ptolemy\inst{2}\fnmsep\thanks{Just to show the %usage
%          of the elements in the author field}
%          }

%   \institute{Institute for Astronomy (IfA), University of Vienna,
%              T\"urkenschanzstrasse 17, A-1180 Vienna\\
%              \email{wuchterl@amok.ast.univie.ac.at}
%         \and
%             University of Alexandria, Department of %Geography, ...\\
%             \email{c.ptolemy@hipparch.uheaven.space}
%             \thanks{The university of heaven temporarily %does not
%                     accept e-mails}
%             }

\author{
Raj Prince\inst{1} \thanks{E-mail: raj@cft.edu.pl}, Aditi Agarwal\inst{2}, Nayantara Gupta\inst{2}, Pratik Majumdar\inst{3}, Bo\.zena Czerny\inst{1} \\
 Sergio A. Cellone\inst{4,5},
I. Andruchow\inst{4,6}}
%Third Author$^{2,3}$
%and Fourth Author$^{3}$

% List of institutions 

\institute{
$^{1}$Center for Theoretical Physics, Polish Academy of Sciences, Al.Lotnikow 32/46, Warsaw, Poland\\
$^{2}$Raman Research Institute, Sadashivanagar, Bangalore 560080, India \\
$^{3}$Saha Institute of Nuclear Physics, HBNI, Kolkata, West Bengal, 700064, India\\
$^{4}$Facultad de Ciencias Astron\'omicas y Geof\'isicas, Universidad Nacional de La Plata, Paseo del Bosque, B1900FWA, La Plata, Argentina\\
$^{5}$Complejo Astron\'omico ``El Leoncito'' (CASLEO), CONICET-UNLP-UNC-UNSJ, San Juan, Argentina \\
$^{6}$Instituto de Astrof\'isica de La Plata (CCT La Plata-CONICET-UNLP), La Plata, Argentina \\
}

  % \date{Received September 15, 1996; accepted March 16, 1997}

% \abstract{}{}{}{}{} 
% 5 {} token are mandatory
 
  \abstract
  {The blazar OJ 287 has been proposed as binary black hole system based on its periodic optical outburst. Among blazars with parsec scale jets, the black hole binary systems are very rare and hence this source is very interesting to study.}
{The BL Lac OJ 287 is an interesting object for multi-wavelength study due to its periodic outbursts. 
We have analysed the optical, X-ray and gamma-ray data of OJ 287 for the period of 2017 -- 2020.
There are several high states in optical-UV and X-ray frequencies during this period. 
Based on the observed variability in optical and X-ray frequencies the entire period 2017 -- 2020 is divided in five segments, referred as A, B, C, D, \& E in this paper. A detailed temporal and spectral analysis is performed to understand the nature of its flaring activities.} 
 {To understand the temporal variability in this source we have studied the intra-day, and fractional variability for all the various states, and along with that fast variability time was also estimated to understand the nature of variability. Further, the multi-wavelength SED modeling is performed to know more about the physical processes responsible for the simultaneous broadband emission and the fast variability.}
{The Fermi LAT observations show a moderate flux level of this source in gamma-ray frequency throughout this period, though flux variability has been observed. The source has shown a strong flux variability in X-ray, optical, and UV during early 2017 and mid 2020 when the source was in very high state. A single zone SSC emission model is considered to model the spectral energy distributions and this helps us to explore the nature of this BL Lac with binary super-massive black holes.}
 %  {Vibrational instability is found to be a common phenomenon at temperatures lower than the second He ionisatio zone. The $\kappa$-mechanism is widespread under `cool' conditions.}
   {}


   \maketitle 
%
%-------------------------------------------------------------------
\section{Introduction}
OJ 287, a BL Lac type active galactic nucleus, located at a redshift of 0.306, was discovered in 1967 (\citealt{Dickel_1967}).  
It was known as an exceptionally active variable source even five decades ago (\citealt{Andrew_1971}). 
A proper study of the variability of OJ 287 in different time scales can be found in \citet{Valtonen_2006}. Intraday variability in radio and optical data of OJ 287 was first detected by \citet{Valtaoja_1985}.
 Variability in blazars fits into three different categories as described below. Changes over a range of minutes to less than a day
\citep[e.g.][]{1995ARA&A..33..163W, 1975IAUS...67..573K, 2003AJ....126...47R} are defined as intra-night variability (INV, or intra-day variability i.e. IDV, or
microvariability); those on a timescale of days to a few months are commonly known as short term variations (STV);
while the variations over several months to years are defined as long term variations \citep[LTV, e.g.][]{2011A&A...531A..38A, 2005A&A...438...39R, 2017MNRAS.469..813A}.

Short time scale variability of this BL Lac object in the near-infrared frequency was studied using standard JHK photometry, which showed variability of amplitude 0.7 mag over the observing period of 23 months (\citealt{Lorenzetti_1989}). From the long-term optical light curve of OJ 287, it was inferred that it has binary supermassive black holes (\citealt{Sillanpaa_1988}). They found that the light curve shows repeated outbursts at the interval of 11.65 years and minimum flux at the interval of 11 years. These results were verified by others (\citealt{Kidger_1992}).
 Different models for the periodic outburst of OJ 287 in optical frequency have been discussed earlier (\citealt{Dey_2019}), involving the periodic motion of a binary supermassive black hole. One kind of model assumes that the orientation of the jet of the primary black hole changes in a regular manner due to precession. The optical flare would thus be the result of the enhancement in the Doppler factor of the jet. In another model, optical flaring in OJ 287 results from enhanced accretion during pericenter passage or collision between the secondary black hole and the accretion disc of the primary black hole.

 A big flare from OJ 287 was predicted to happen in 1994 according to the binary black hole model of \citet{Sillanpaa_1988}. This was observed by \citet{Sillanpaa_1996} and thus the prediction of 12 year cycle was confirmed. \citet{Lehto_1996} proposed that the reason for the flares is an impact of the secondary black hole on the accretion disk of the primary, which means that there has to be two such flares during each orbital cycle. This is a unique property of the model, not easily accounted for in other proposals. The model predicted the time of the second flare in November 1995 within a two week time window (\citealt{Valtonen_1996}), and subsequently \citet{Sillanpaa_1996b} observed the flare and confirmed the prediction. \citet{Sundelius_1996, Sundelius_1997} calculated the flare arising from tides in this binary model and predicted the next big impact flare in 2005, a year earlier than would be expected from strict periodicity. It was reported by \citet{Valtonen_2006}. The flares come sooner than in the strictly periodic models due to precession, as is well stated in their paper. Finally, the observation of the 2015 flare confirmed this shift, which by then was 3 years (\citealt{Valtonen_2016}). This paper also found the signature of disk impacts, the thermal nature of the flare. \citet{2019ApJ...882...88V} updated the model of \citet{Lehto_1996} and determined the disk parameters using time delays calculated in \citet{Dey_2018}. 
\par
 During the phase 2008 – 2010 tidal flares were expected according to the model by \citet{Sundelius_1996,Sundelius_1997}.
The gamma-ray light curve of OJ 287 during 2008 August -- 2010 January was studied by \citet{Neronov_2011}. They found the variability time scale is lower than 3.2 hours. They inferred that the observed gamma-ray emission was from the jet of the smaller mass black hole. Detection of gamma-rays of energy higher than 10 GeV constrained the lower limit of the Doppler factor to 4.
\par
The broadband spectrum of the major gamma-ray flare in 2009 was studied by \citet{Kushwaha_2013}. They explained the multi-wavelength spectral energy distribution (SED) by combining synchrotron, synchrotron self-Compton (SSC), and external Compton (EC) processes. They suggested that the emission region in the jet is surrounded by a bath of photons at 250 K. They also inferred that the location of this emission region is 9 pc away from the central engine. The high activity of OJ 287 during December 2015 -- April 2016 was studied by \citet{2018MNRAS.473.1145K}, and the authors inferred simultaneous multi-wavelength emission. They explained the optical bump as accretion disc emission associated with the primary black hole.
 The smaller bump feature in optical-UV appeared to be consistent with line emission. They explained the gamma-ray emission with inverse Compton scattering of photons from the line emission.
 \par
 The flux and polarisation variability at optical bands of OJ 287 during the December 2015 to February 2016 outburst was studied by \citet{Rakshit_2017}. The intra-night optical variability data was analyzed, and the shortest variability time scale was estimated as $142\pm 38$ minutes. This constrained the lower limit on the value of the Doppler factor to 1.17 and the upper limit on the value of the magnetic field to 3.8 Gauss. The size of the emission region was constrained to less than $2.28\times 10^{14}$ cm.
 \par
 The multi-band optical variability from September 2015 to May 2016 was studied by \citet{Gupta_2017} using nine ground-based optical telescopes. They detected a large optical outburst in December 2015 and a second comparably strong flare in March 2016. The long term optical, ultraviolet and X-ray variability in different activity states of OJ 287 was studied using 
 UVOT and XRT instruments of Swift (\citealt{Siejkowski_2017}). They did not find any clear relation between optical/UV and X-ray emission during quiescence state or outbursts.
 \par
 The strong activity in optical to X-ray frequency during July 2016 to July 2017 was studied by \citet{2018MNRAS.479.1672K}. The daily gamma-ray fluxes during this time are consistent with no variability. They modeled the SEDs with a two-zone leptonic model. The first zone gives an LBL SED, and the second zone gives an HBL SED. In their model, the second zone is located at a distance of parsec-scale from the central engine. 
 \par
 A hadronic model to explain the X-ray and gamma-ray outburst of November 2015 outburst of OJ 287 has been given by \citet{2020MNRAS.498.5424R}. They have used the binary supermassive black hole model, where the initial trigger comes from the impact of the secondary black hole on the accretion disc of the primary black hole. An idealized spherical outflow is generated from this impact. A shock is formed when this spherical outflow -- containing cosmic rays and thermal ions -- interacts with the AGN wind of the primary black hole.
 In their model, the cosmic rays are shock accelerated due to the collision of the outflow with the AGN wind of the primary black hole. The cosmic ray protons interact with the thermal ions, and as a result, secondary leptons, photons are produced in proton-proton interactions. The optical flare is explained by combining the jet emission from \citet{Kushwaha_2013} and the thermal bremsstrahlung emission in the outflow.
 The photon field produced as a result of thermal bremsstrahlung acts as target for inverse Compton emission by the secondary leptons. They have explained the X-ray and gamma-ray data by this inverse Compton emission of the secondary electrons. 
 \par
 Recently, \citet{Komossa_2020} reported the detection of a very bright outburst of OJ 287 covering X-ray, UV, and optical frequency from April to June of 2020. %This is the second brightest outburst since they started multi-layer monitoring of this source with Swift in late 2015. 
 They concluded that the outburst is jet-driven and consistent with the binary supermassive black hole model. In this model, the impact of the secondary black hole on the disk of the primary triggers an after-flare. This impact enhances the accretion activity of the primary black hole, which results in enhanced jet emission by the primary black hole.
 \par
 In this paper, we have analyzed the multi-wavelength data of OJ 287 for the period of 2017 to 2020, which includes the outburst discussed in the paper by \citet{Komossa_2020}. The total period 2017 -- 2020 considered in our work has been divided into five segments after analyzing the variability time scale in optical and X-ray data. We have done the modeling of the SEDs with a time-dependent leptonic model, which includes synchrotron, SSC. 
  The data analysis is discussed in section 2. Our results of data analysis 
  and the modelling of SEDs are discussed in section 3.
 The discussions and conclusions of our study are given in section 4. 

\section{Multiwavelength Observations and Data Analysis}

\subsection{\textit{Fermi}-LAT}
\textit{Fermi-LAT} is an excellent space-based telescope to explore the extragalactic and Galactic objects in the gamma-ray sky. It uses the pair conversion method to detect gamma-rays in the energy
range of 20 MeV -- 500 GeV. It has a wide field of view (FoV) of about 2.4 sr (\citealt{Atwood_2009}), which scans 20\% of the sky at any time.
The total scanning period of the entire sky with this telescope is around three hours. OJ 287 was observed in the brightest flaring state in X-ray when monitored by the Swift-telescope (Atel 10043) in 2017, and, soon, flares in other frequency bands were also detected.
\textit{Fermi}-LAT is continuously monitoring the source OJ 287 since 2008. We have collected the data from January 2017 to May 2020, and it is found that the source is in a moderate flux state within this period. 
 We have analyzed the gamma-ray data following the standard data reduction and analysis procedure described by 
\textit{Science Tools}\footnote{https://fermi.gsfc.nasa.gov/ssc/data/analysis/documentation/}. 
The details of the method of this analysis are discussed in \citet{Prince_2018}.
 
\begin{figure*}
%\vspace{-10pt}
 \centering
 \includegraphics[scale=0.60]{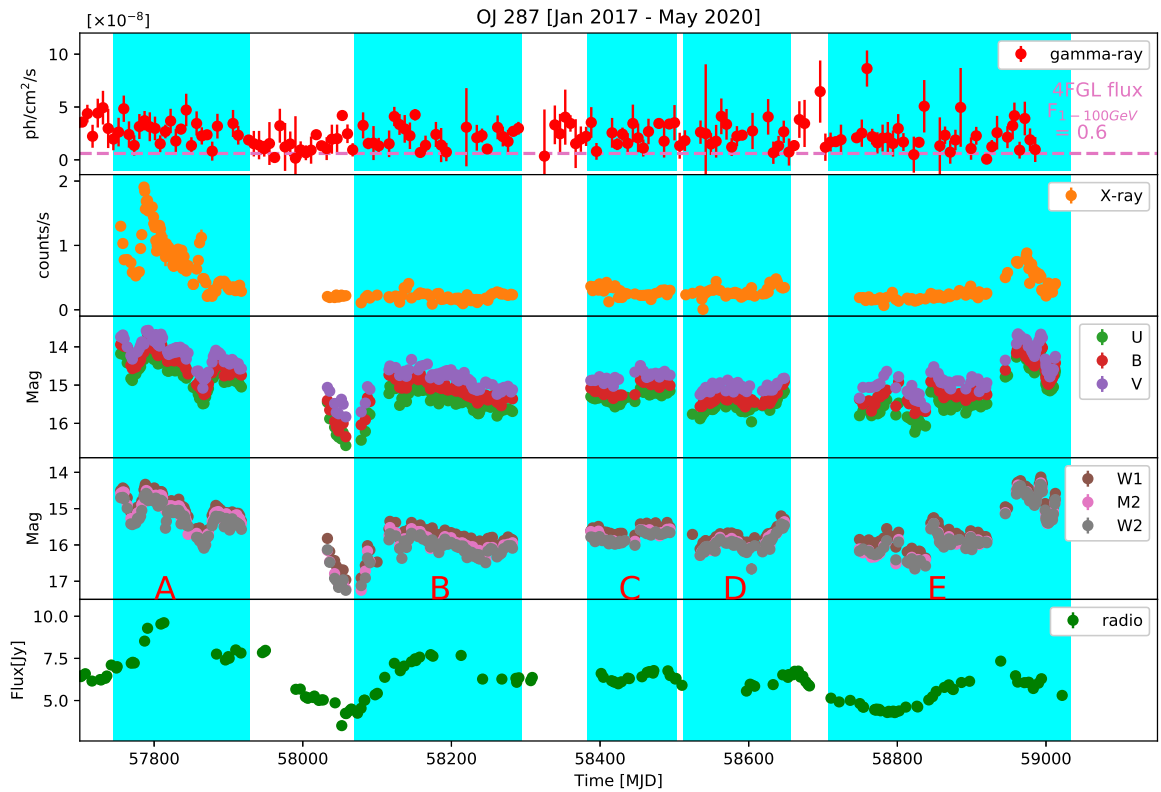}
 %\includegraphics[scale=0.45]{total_1day.eps}
 %\vspace{-50pt}
 \caption{The upper plot shows the weekly binned gamma-ray light curve for 0.1--300 GeV. Panels 2nd, 3rd, and 4th are the Swift-XRT and UVOT light curves. The 5th panel is the radio light curve from OVRO at 15 GHz.
 The entire light curve is divided in five different states based on the flux and magnitude seen in Swift-XRT and UVOT. The various states are denoted as A, B, C, D, and E and their time duration is represented by the color patches.}
 \label{fig:total_lc}
\end{figure*}

\subsection{X-ray Observations}
On February 3, 2017, an X-ray flare was observed by the \textit{Swift}-telescope, and the results were reported in Atel 10043. It is reported as the brightest flare ever detected since the monitoring started by the \textit{Swift}-telescope. After that, many multiple flares were observed in X-rays until May 2020, and this whole period has been studied in this paper.
\textit{Swift} is a space-based telescope with three instruments onboard, observing all kinds of Galactic and extragalactic sources in soft $\&$ hard X-rays, Optical, and UV simultaneously. The working energy range of Swift-XRT is 0.3-10.0 keV.
The BL Lac OJ 287 was observed by \textit{Swift-XRT} telescope during the multiple flaring episodes in X-ray frequencies between the period January 2017 to May 2020. We have analyzed all the observations done during this period, and the processing of the raw data is done by using the task `\textit{xrtpipeline}\footnote{https://heasarc.gsfc.nasa.gov/ftools/caldb/help/xrtpipeline.html}' and cleaned events file are produced for each observation. 
The CALDB version 20160609 is used while processing the raw data. Our analysis is focused on only the Photon Counting
mode observations, and the task `\textit{xselect}' is used for source and background selection. We have selected a region of 12 arc seconds around
the source and away from the source for the source and background, respectively, in our data analysis. The task `\textit{xselect}' is also used to extract the spectrum and light curve, and the modeling
of the spectrum is done in `\textit{Xspec}'(\citealt{Arnaud_1996}). For modeling the spectra, we have used a single power-law model. The Galactic absorption column
density $n_H$ = 1.10$\times$10$^{20}$ cm$^{-2}$ is fixed from \citet{Kalberla_2005}. The modeling is done for an energy range of 0.3 -- 10.0 keV.

\subsection{Optical and UV Observations}
Having the Swift Ultraviolet/Optical Telescope (UVOT, \citealt{Roming_2005}) on board with \textit{Swift-XRT} has the advantage of getting simultaneous observations in Optical and UV bands. 
\textit{Swift-UVOT} has also observed the OJ 287 in all of the available six filters U, V, B, W1, M2, and W2, simultaneously with the X-ray observations. 
The source instrumental magnitudes are extracted following the $uvotsource$ procedure. We have considered the region of 5 arcsec around the source and away from it as the source and the background region, respectively, in our data analysis.  
The magnitudes are corrected for galactic extinction by using the reddening E(B-V) = 0.0241 from \citet{Schlafly_2011} and zero points from \citet{Breeveld_2011}. Moreover, the magnitudes are converted into flux by multiplying by the conversion factor estimated by \citet{Poole_2008} and the ratios of extinction to reddening from \citet{Giommi_2006}. 

 In the period between Feb 2019 - Jan 2020, we performed observations of OJ 287 using five different telescopes around the globe, which are: 2.15\,m Jorge Sahade telescope (JS, telescope A) and 60\, cm Helen Sawyer Hogg telescope (HSH, telescope B), CASLEO, Argentina;
1.3\,m JC Bhattacharya telescope (JCBT; telescope C) at the Vainu Bappu Observatory (VBO), India.
The technical descriptions of the above telescopes are summarized in Table 1 of \citet{2019MNRAS.488.4093A} and
\citet{2021A&A...645A.137A}. The number of observations made in each band on a particular date during our monitoring campaign
is provided in Table\,\ref{tab:obs_log}.

The preliminary data reduction includes bias correction, flat fielding, and cosmic-ray removal, which was performed with
IRAF\footnote{IRAF is distributed by the National Optical Astronomy Observatories, which are operated
by the Association of Universities for Research in Astronomy Inc., under a cooperative agreement with the
National Science Foundation.} software. We then processed the cleaned CCD images using the Dominion Astronomical
Observatory Photometry (DAOPHOT II) software \citep{S1987PASP, S1992ASPC} using the aperture photometry
technique through which we obtained instrumental magnitudes for our target and four standard stars located in the same field. A more detailed and comprehensive description of data reduction methods used is given in Section 2 of \citet{2019MNRAS.488.4093A}. Finally, to extract the instrumental
differential light curves (LCs), we selected two non-variable standards having
magnitude and color very similar to that of the blazar. The calibrated LCs were obtained using the star 10 of \citet{1996A&AS..116..403F}.
After constructing the calibrated LCs of our source, we carefully inspected the LCs for any outliers. A handful of
such suspicious data points were detected and corrected.

\subsection{Radio data at 15 GHz}
Owens Valley Radio Observatory(OVRO; \citealt{Richards_2011}) is one of the observatories that monitors the bright \textit{Fermi} detected blazars. It is a 40-meter single-dish antenna working at a frequency of 15 GHz. A large number of Fermi blazars are continuously monitored by OVRO twice a week. Our candidate source, OJ 287, is also part of the OVRO monitoring program, and we have collected the data from September 2017 to July 2020. 

\begin{table} 
\caption{Log of photometric observations for the blazar OJ\,287. }
\textwidth=8.0in
\textheight=11.0in
 \centering
\noindent
\label{tab:obs_log} 
\begin{tabular}{lc|llll} \\ %{lclllll}
\multicolumn{2}{c}{}\\
\hline 
\noalign{\smallskip} 
  Date of        & Telescope      & \multicolumn{4}{c}{Number of data points}  \\
observations    &     &     \\
(yyyy mm dd)     &  & $B$ & $V$ & $R$ & $I$      \\
\noalign{\smallskip} 
\hline 
\noalign{\smallskip} 
2019 02 15    & C  & 0 & 1 & 6 & 1 \\
2019 02 27    & A  & 3 & 2 & 30 & 2  \\
 2019 03 01    & A  &2 & 9 & 27 & 1  \\
2019 03 02    & A  &0 & 9 & 22 & 1  \\
2019 03 03    & A  &1 &17 &16 & 2  \\
2019 03 09    & C  &1 & 1 & 1 & 1  \\
2019 03 11    & C  &1 & 1 &1 & 1  \\
2019 03 12    & C  &0 & 1 &1 & 0  \\
 2019 03 13    & C  &1 & 1 & 1 & 1  \\
 2019 03 26    & A  &2 & 1 &24& 2 \\
 2019 04 05    & C  &1&1 &17 &1 \\
 2019 04 06    & C  &1 & 1 & 10 & 1  \\
 2019 04 07    & C  &1 & 2 &20 & 2  \\
 2019 04 09    & A  &1 &16 &16 &1  \\
 2019 04 10    & A  &2 & 10 &10 & 2  \\
 2019 04 10    & A  &3 & 4 & 12 & 2  \\
 2019 04 11    & A  &2 & 12 &14 & 2  \\
 2019 12 17    & A  &0 & 1 &9 & 2  \\
 2020 01 03    & A  &2 & 2 &120 & 2  \\
 2020 01 27    & B &0 & 4 &4 & 1  \\ 
 \noalign{\smallskip} 
  \hline
  \end{tabular} 
\end{table}

\begin{table*}
 \centering
\caption{Table shows the fractional variability and the variability time estimated for various states in different waveband as shown in Figure 1, and explained in section 3.1.1 in detail.}
 \begin{tabular}{ccc c p{0.1cm}}
  \hline
  \noalign{\smallskip} 
 Instrument& Various states& Fractional variability& Variability time\\
           &    &F$_{\rm var}$ & $\tau_{var}$[days] \\ 
           \noalign{\smallskip} 
\hline 
\noalign{\smallskip} 
  XRT      & A     & 0.48$\pm$0.01   &   2.41 &      \\  
  XRT      & B     & 0.29$\pm$0.01   &   2.39 &    \\
  XRT      & C     & 0.25$\pm$0.01   &   2.32 &    \\
  XRT      & D     & 0.28$\pm$0.01   &   0.80 &    \\
  XRT      & E     & 0.57$\pm$0.01   &   0.98  &    \\ 
%  All      &   &   &                   &       &   \\
\noalign{\smallskip} 
  \hline
  \noalign{\smallskip} 
 UVOT-U &   A   &0.284$\pm$0.003 &   4.99&  \\
 U      &   B   &0.215$\pm$0.004 &  3.19&   \\
 U      &   C   &0.103$\pm$0.006 &  15.83   \\
 U     &   D    &0.164$\pm$0.005 &  6.94    \\
 U     &   E    &0.585$\pm$0.003  &  0.58    \\ 
% All   &       &   &                &          \\
\noalign{\smallskip}
 \hline
 \noalign{\smallskip}
 UVOT-B &    A   &0.191$\pm$0.032 &  3.75    \\
 B      &    B   &0.221$\pm$0.004 &  3.18    \\
 B      &    C   &0.140$\pm$0.006 &  19.70   \\
 B      &    D   &0.152$\pm$0.005 &  3.56    \\
 B      &    E   &0.483$\pm$0.003 &  1.26    \\ 
% All    &   &       &   &                &          \\
 \noalign{\smallskip}
 \hline
 \noalign{\smallskip}
 UVOT-V &  A   &0.272$\pm$0.003 &  5.25    \\
 V      &   B   &0.228$\pm$0.005 &  4.04    \\
 V      &   C   &0.104$\pm$0.008 &  10.33   \\
 V      &   D   &0.088$\pm$0.007 &  2.82    \\
 V      &   E   &0.499$\pm$0.004 &  0.76    \\ 
% All    &   &       &   &                &          \\
 \noalign{\smallskip}
 \hline 
 \noalign{\smallskip}
 UVOT-W1&   A   &0.303$\pm$0.003 &  5.44    \\
 W1     &   B   &0.205$\pm$0.005 &  5.97    \\
 W1     &   C   &0.096$\pm$0.007 &  11.26   \\
 W1     &   D   &0.203$\pm$0.006 &  6.86    \\
 W1     &   E   &0.599$\pm$0.004 &  0.95    \\ 
% All    &   &       &   &                &          \\
\noalign{\smallskip}
\hline 
\noalign{\smallskip}
 UVOT-M2&   A   &0.305$\pm$0.001 &  4.00    \\
 M2     &   B   &0.222$\pm$0.002 &  5.06    \\
 M2     &   C   &0.117$\pm$0.003 &  19.67   \\
 M2     &   D   &0.205$\pm$0.003 &  6.53    \\
 M2     &   E   &0.635$\pm$0.002 &  3.80    \\ 
% All    &   &       &   &                &          \\
\noalign{\smallskip}
\hline 
\noalign{\smallskip}
 UVOT-W2&   A   &0.308$\pm$0.003 &  4.15   \\
 W2     &   B   &0.214$\pm$0.004 &  4.60    \\
 W2     &   C   &0.111$\pm$0.006 &  19.28   \\
 W2     &   D   &0.221$\pm$0.006 &  3.53    \\
 W2     &   E   &0.628$\pm$0.004 &  1.08    \\ 
 \noalign{\smallskip}
% All    &   &       &   &                &          \\
 \hline 
  \end{tabular}
  \label{tab:var}
\end{table*}

\begin{table*}
\caption{Results of INV observations of OJ 287.} 
\label{tab:var_res}
\centering 
\resizebox{0.82\textwidth}{!}{ 
\begin{tabular}{ccccccccc} 
\hline\hline \noalign{\smallskip}
Date of observation & Passband & $N$ & $\sigma_1$ & $\sigma_2$ & $\Gamma_{\rm SF}$ & $C$-test & $F$-test & Variable (?)\\ \noalign{\smallskip}
 (yyyy mm dd)       &          &     &            &            &          &          &          & \\ \noalign{\smallskip} 
 (1)                &   (2)    & (3) & (4)        &  (5)       & (6)      & (7)      & (8)      & (9)\\
 \noalign{\smallskip} 
 \hline
 \noalign{\smallskip} 
 27.02.2019        &   R      &  30  &  0.0063   & 0.0027     &  1.1219   & 2.3389  & 5.4705    & PV \\
 01.03.2019        &   R      &  27  &  0.0105   & 0.0041     &  1.1073   & 2.5505  & 6.5051    & PV \\
 02.03.2019        &   R      &  22  &  0.0031   & 0.0037     &  1.1287   & 0.8236  & 1.4741    & NV \\
 03.03.2019        &   R      &  16  &  0.0026   & 0.0033     &  1.1166   & 0.7715  & 1.6802    & NV \\
                   &   V      &  17  &  0.0143   & 0.0075     & 1.1177    & 1.8953  & 3.5921    & NV \\
 26.03.2019        &   R      &  24  &  0.0064   & 0.0027     & 1.0808    & 2.3377  & 5.4647    & PV \\         
 05.04.2019        &   R      &  17  &  0.0117   & 0.0133     & 1.1345    & 0.8754  & 1.3049    & NV \\
 06.04.2019        &   R      &  17  &  0.0139   & 0.0128     & 1.1406    & 1.0883  & 1.1845    & NV \\
 07.04.2019        &   R      &  20  &  0.0289   & 0.0088     & 1.0617    & 3.2657  & 10.665    & Var\\
 09.04.2019        &   R      &  16  &  0.0119   & 0.0063     & 1.0389    & 1.8821  & 3.5423    & NV \\
                   &   V      &  16  &  0.0182   & 0.0068     & 1.0549    & 2.6690  & 7.1238    & Var\\
 10.04.2019        &   R      &  10  &  0.0052   &  0.0055    & 1.0721    & 0.9523  & 1.1027    & NV \\
                   &   V      &  10  &  0.0068   & 0.0024     & 1.0890    & 2.8971  & 8.3932    & Var\\
 10.04.2019        &   R      &  12  &  0.0058   & 0.0049     & 1.0980    & 1.1759  & 1.3828    & NV \\
 11.04.2019        &   R      &  14  &  0.0086   & 0.0042     & 1.0969    & 2.0671  & 4.2730    & PV \\   
                   &   V      &  12  &  0.0140   & 0.0035     & 1.0994    & 3.9829  & 15.863    & Var\\
 03.01.2020        &   R      &  120 &  0.0076   & 0.0050     & 0.9357    & 1.5211  & 2.3137    & PV \\  

 \noalign{\smallskip} \hline\noalign{\smallskip} 
 \end{tabular}}\\
\tablefoot{Table columns read: (2) passband of observation. (3) Number of data points in the given passband. (4)-(5) Results for $C$ and $F$-test, respectively. (6) Corresponding scale factor. (7) Dispersion of the corresponding control-comparison star LC. (8) Variability status denoted as follows: Var = variable, NV = non-variable, PV = possibly variable.}
\end{table*}

\begin{figure*}
\begin{center}
\epsfig{figure=  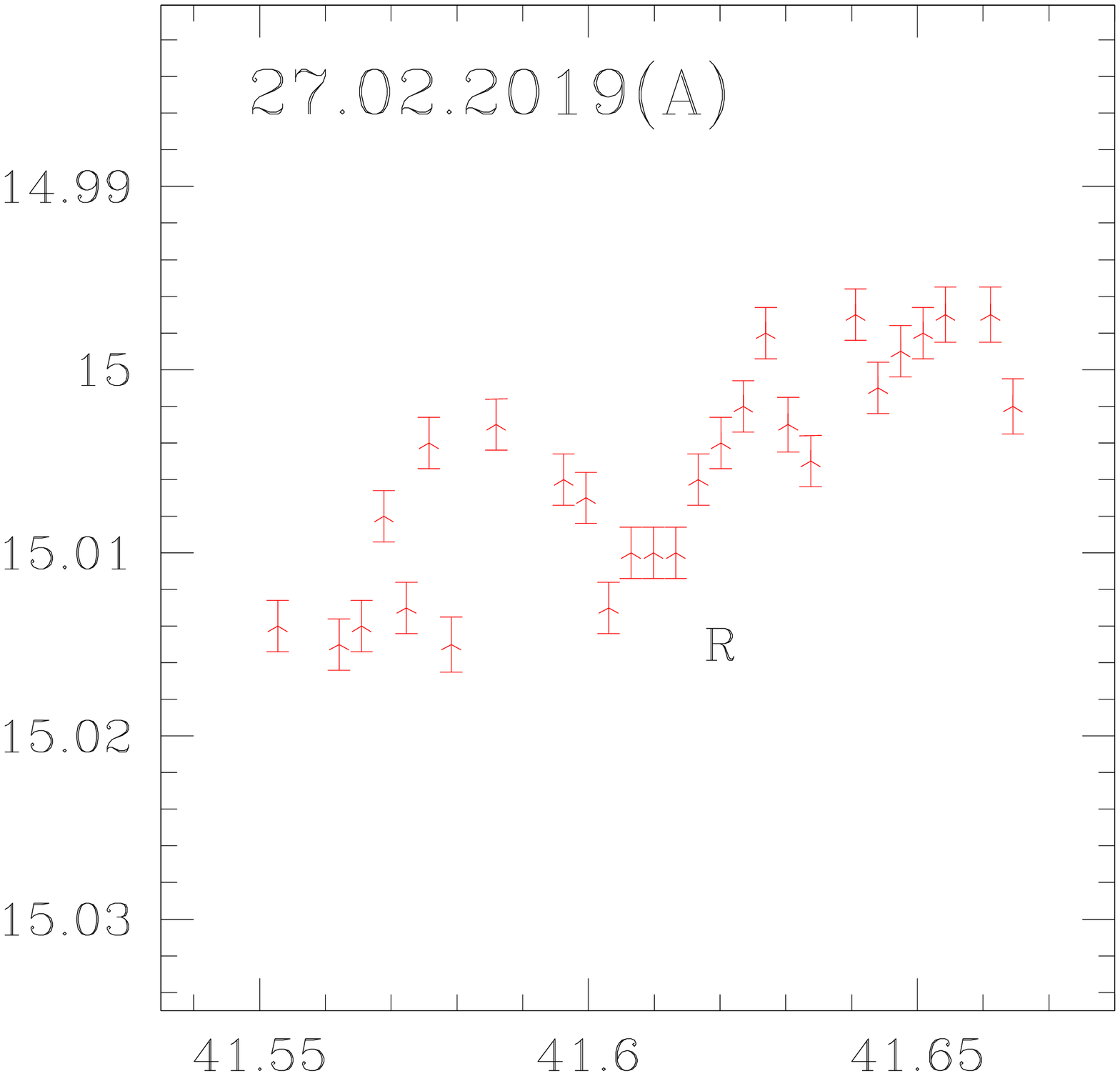,height=1.8in,width=2in,angle=0}
\epsfig{figure=  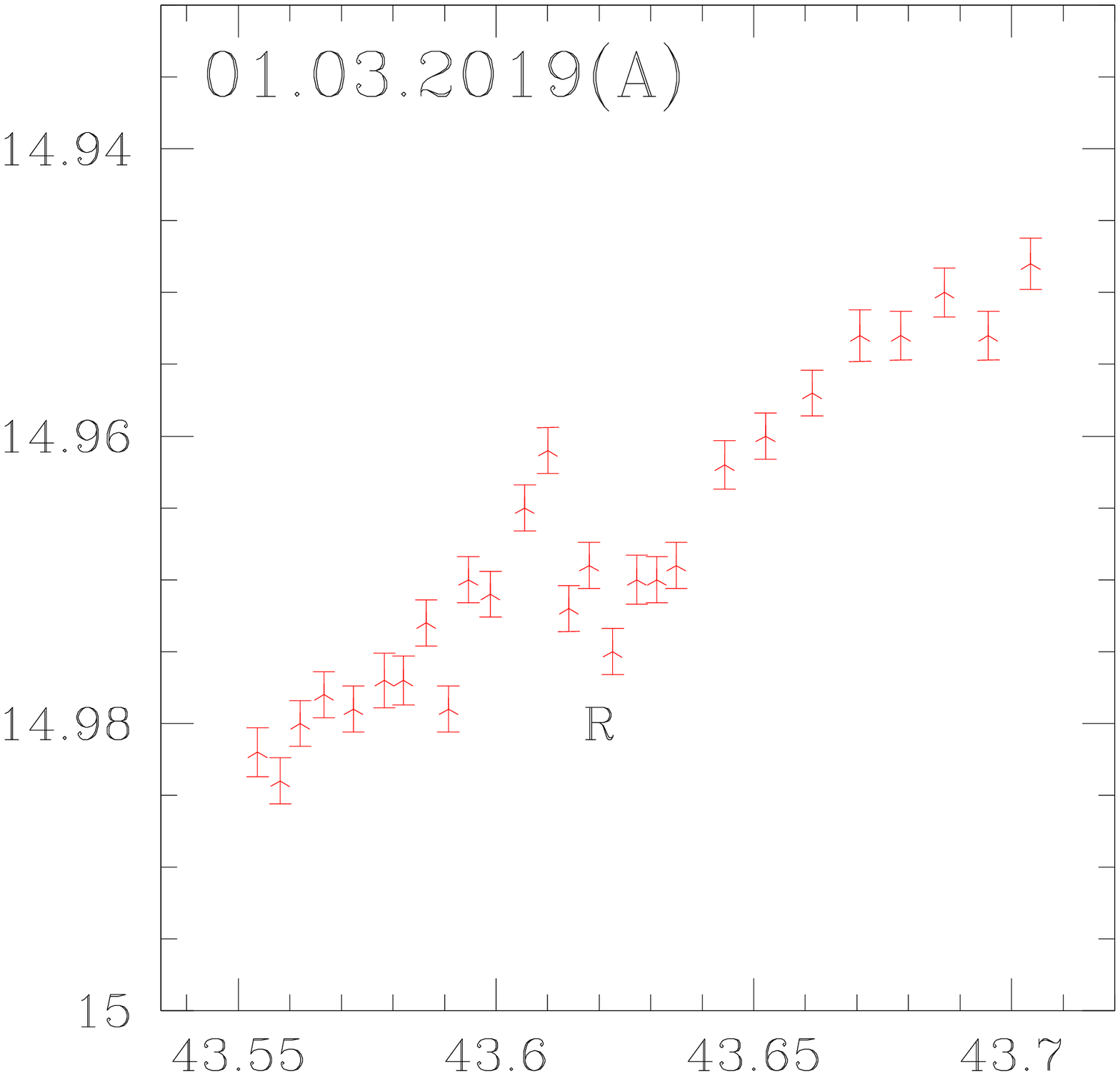,height=1.8in,width=2in,angle=0}
\epsfig{figure=  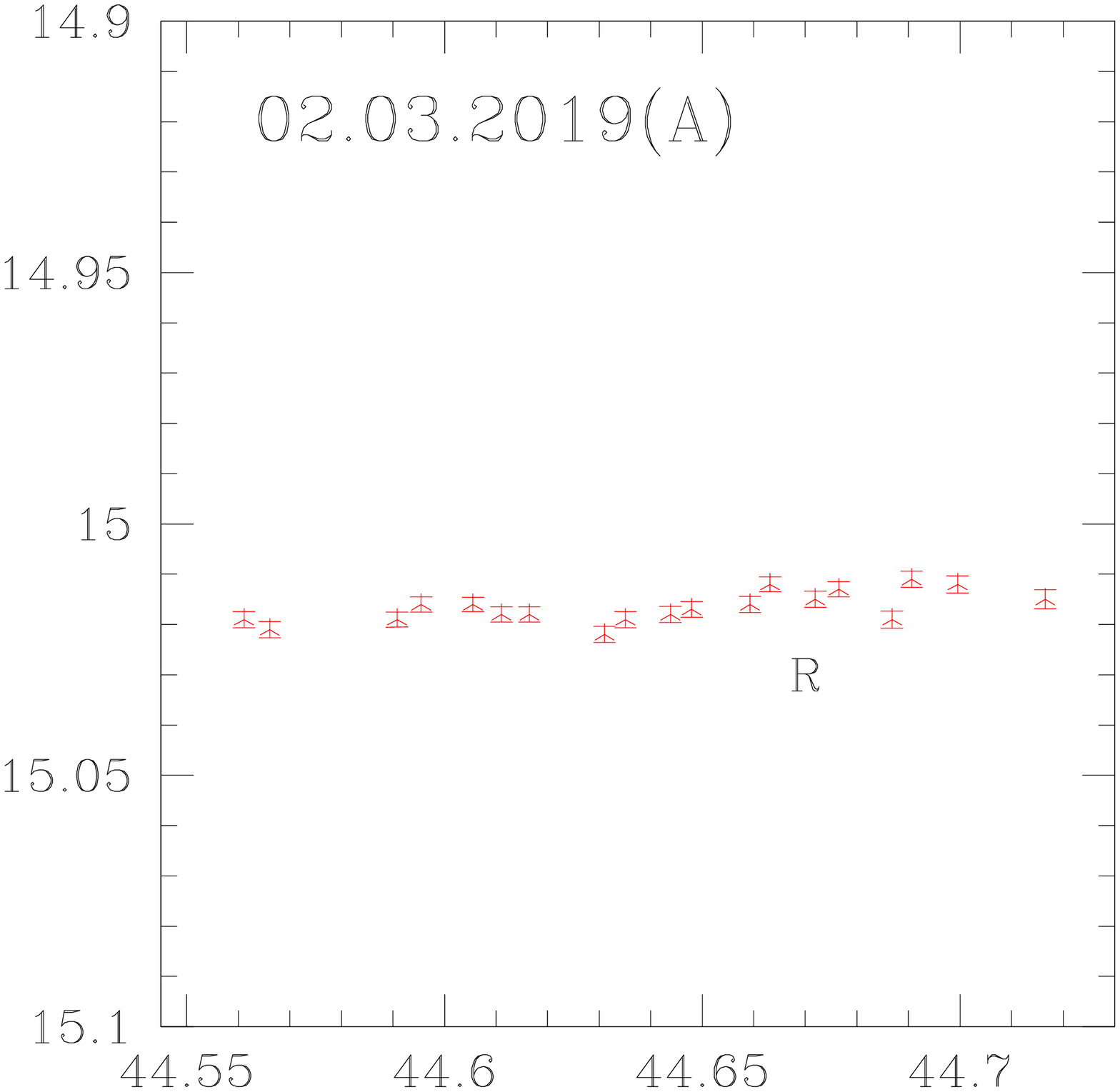,height=1.8in,width=2in,angle=0}
\epsfig{figure=  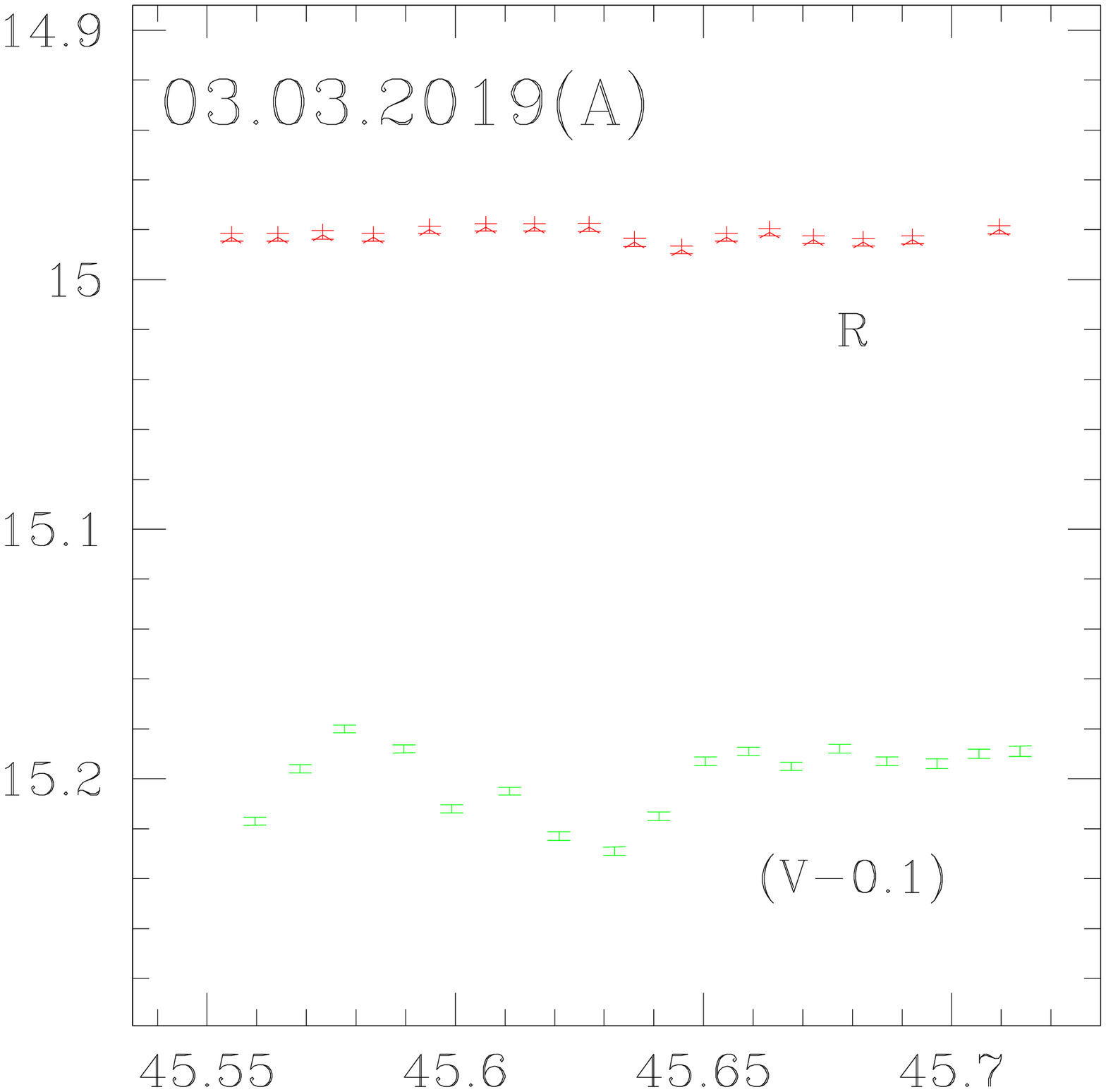,height=1.8in,width=2in,angle=0}
\epsfig{figure=  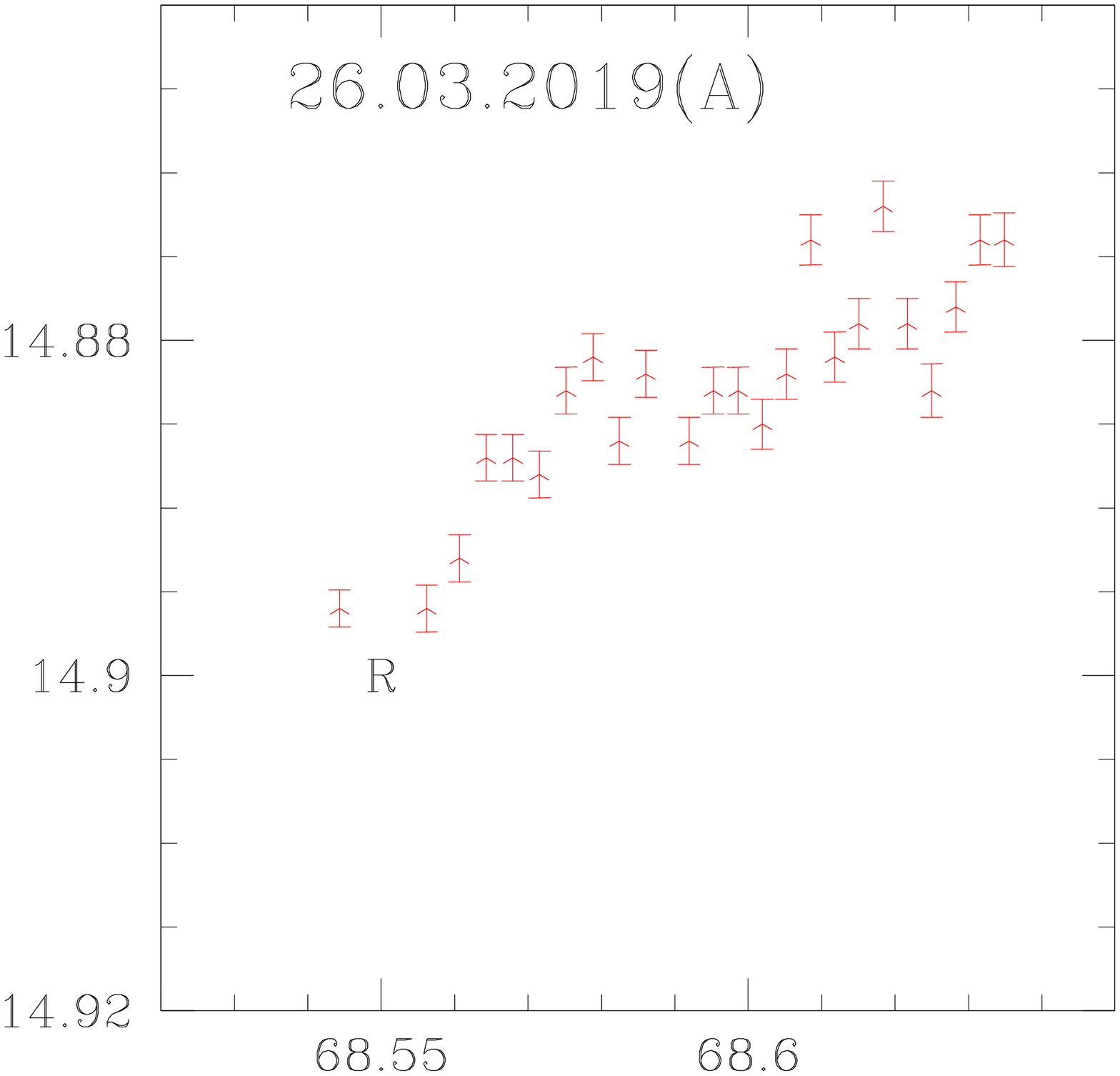,height=1.8in,width=2in,angle=0}
\epsfig{figure=  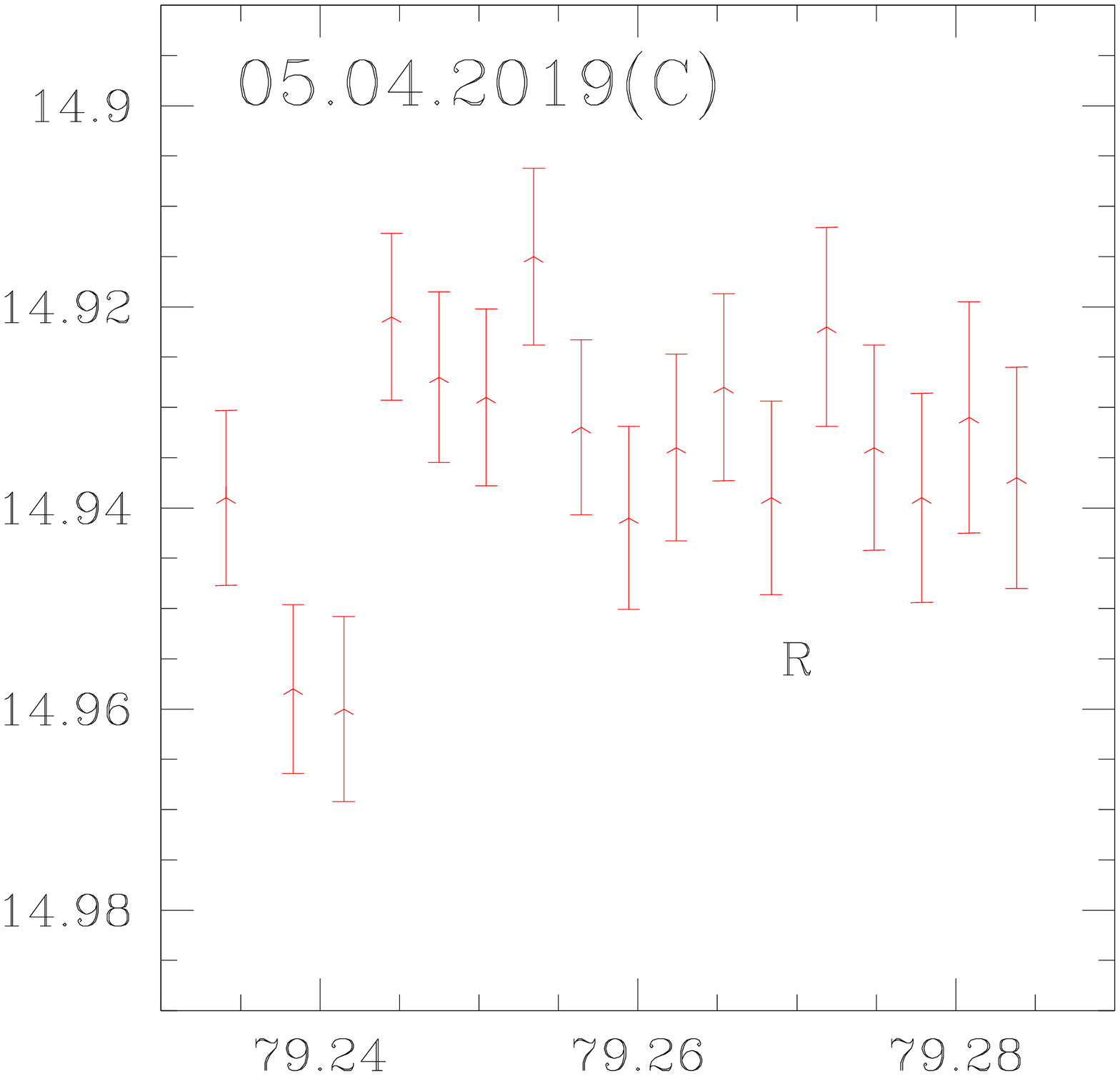,height=1.8in,width=2in,angle=0}
\epsfig{figure=  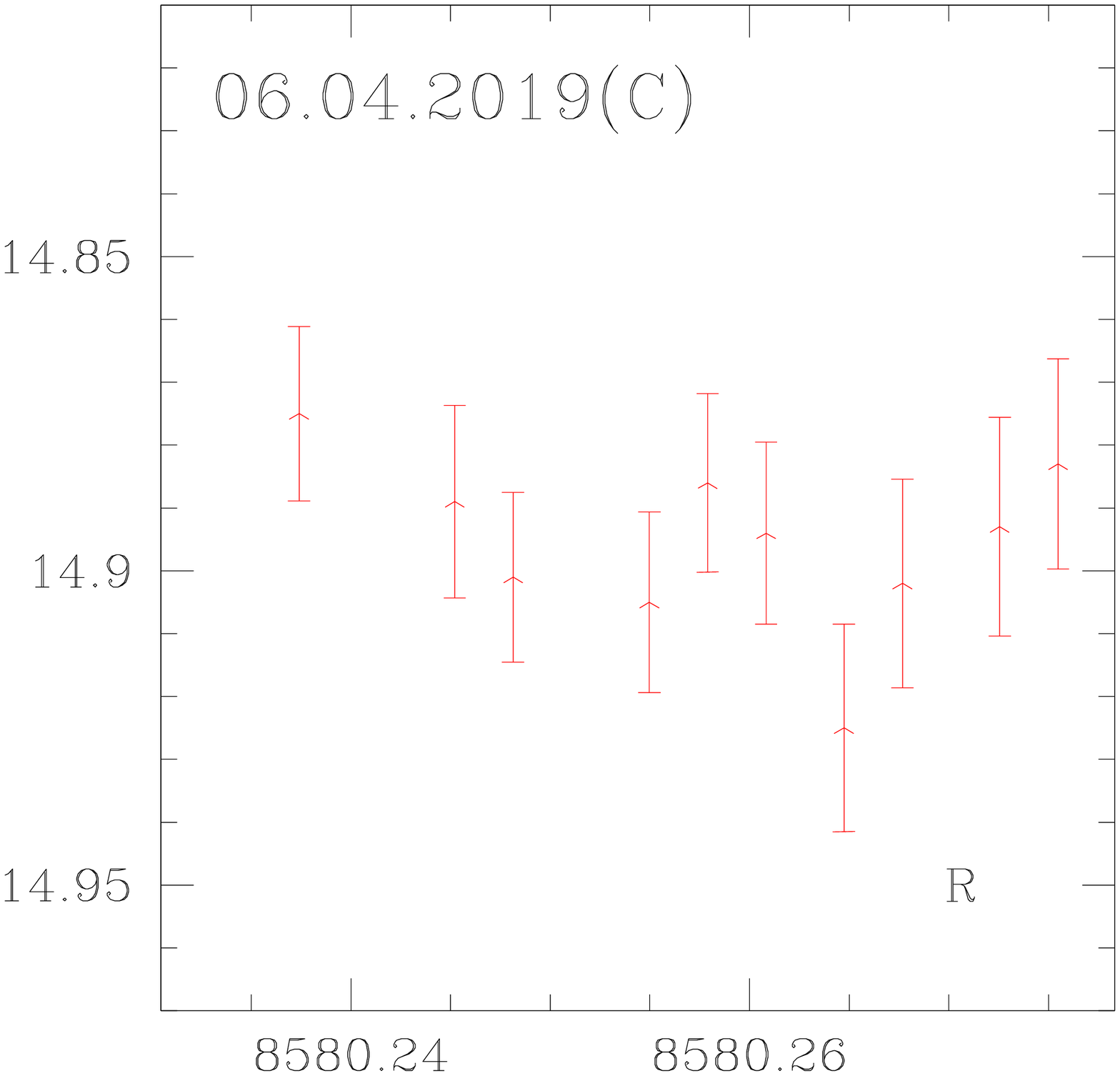,height=1.8in,width=2in,angle=0}
\epsfig{figure=  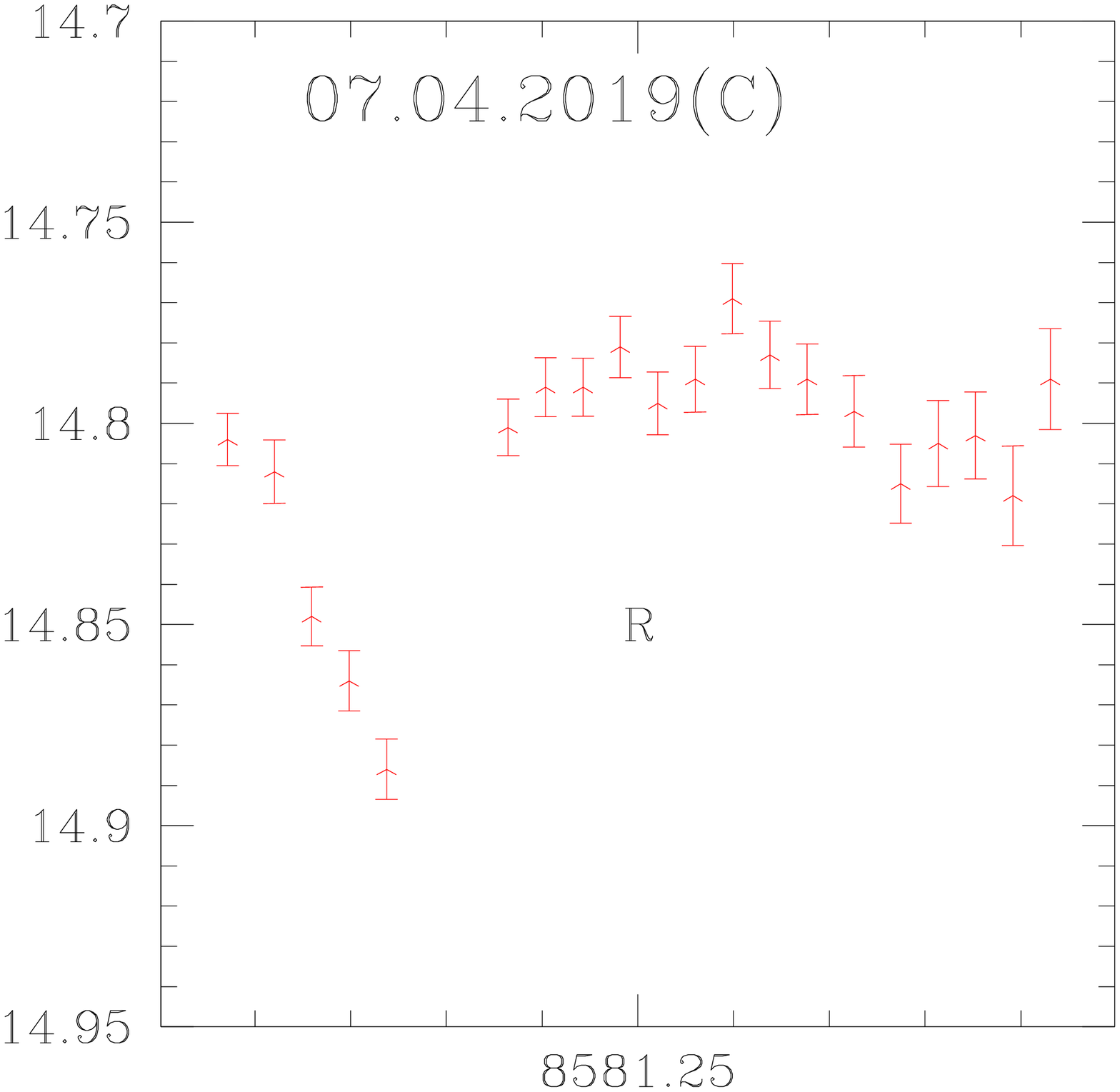,height=1.8in,width=2in,angle=0}
\epsfig{figure=  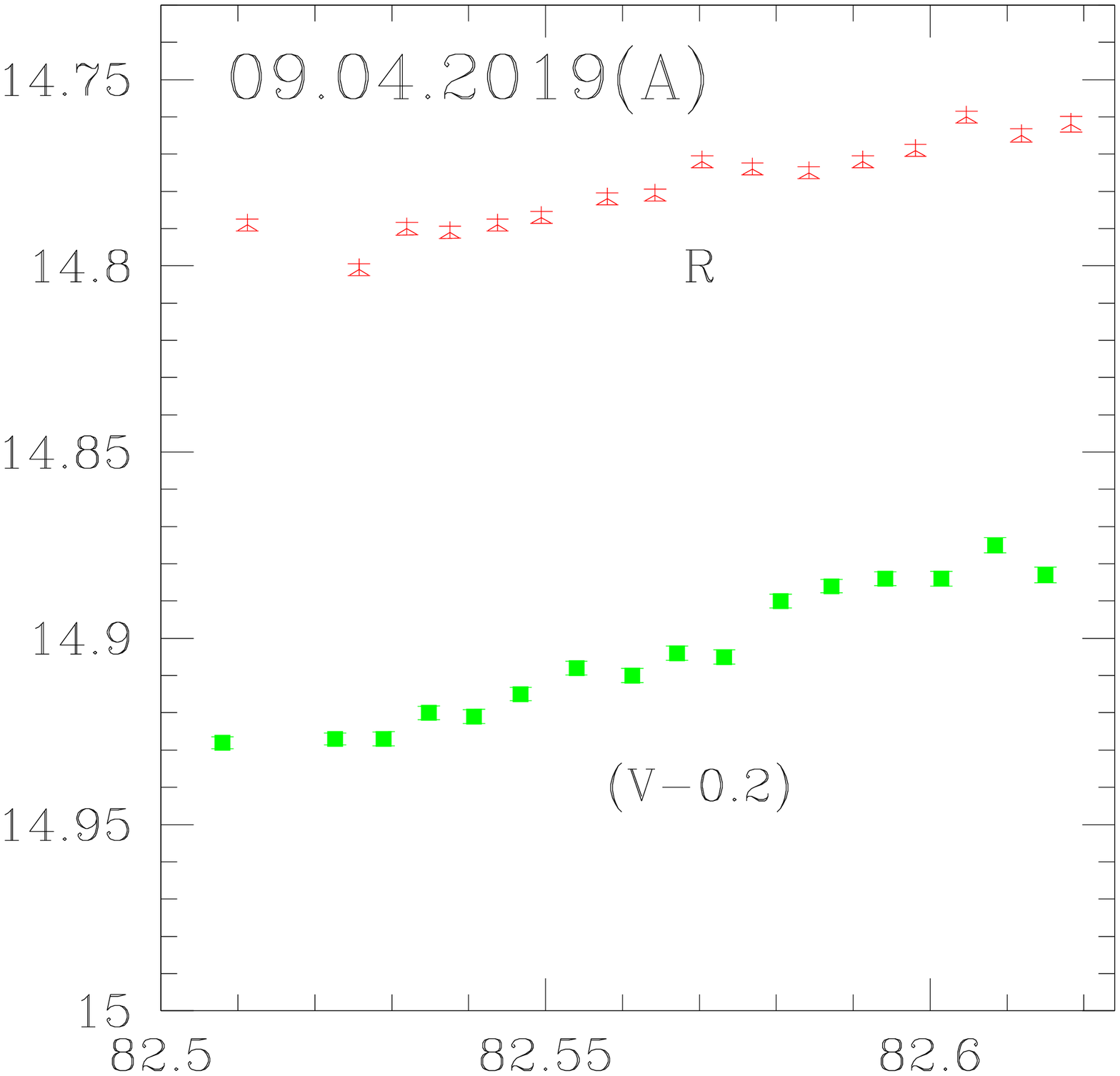,height=1.8in,width=2in,angle=0}
\epsfig{figure=  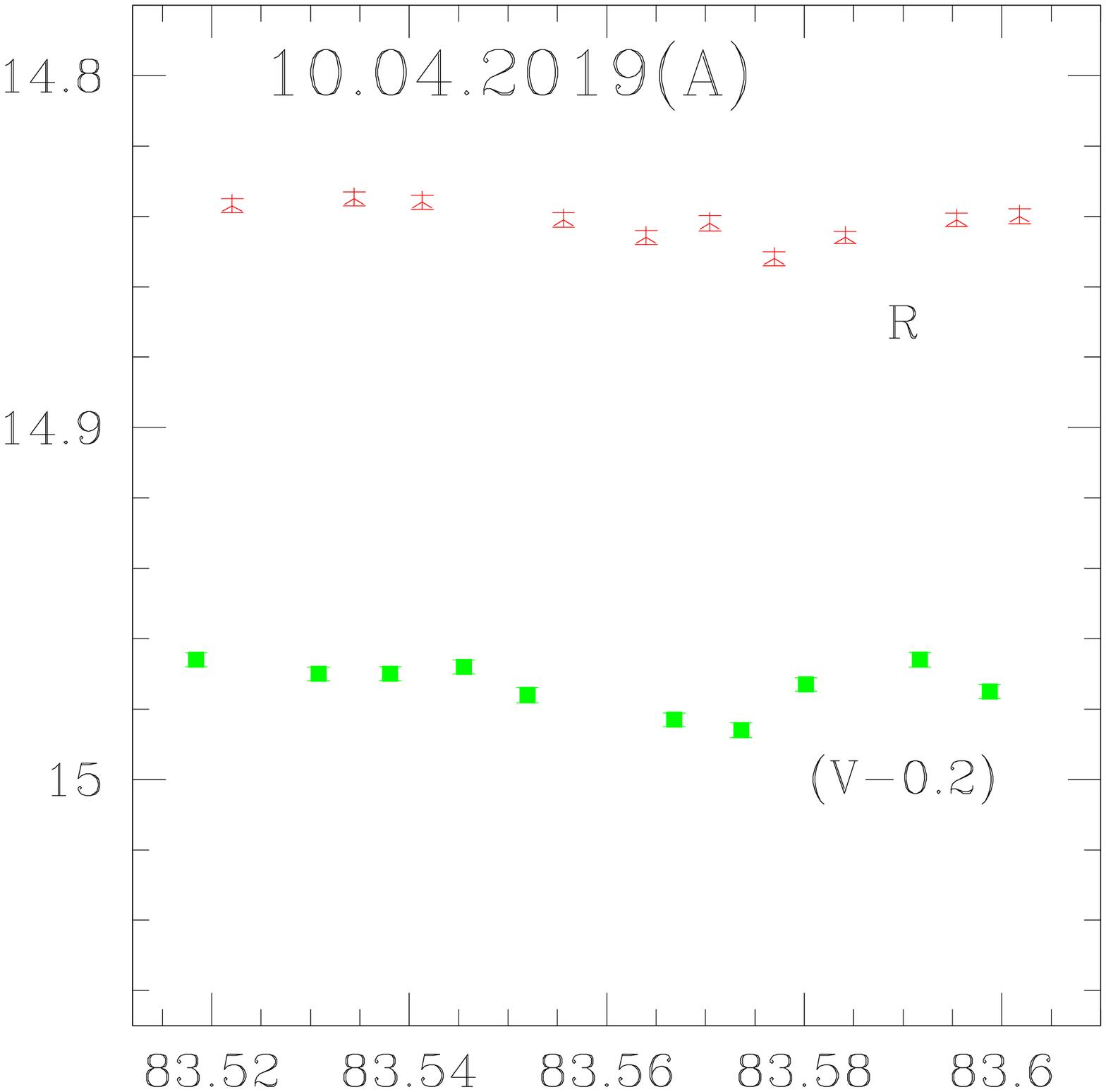,height=1.8in,width=2in,angle=0}
\epsfig{figure=  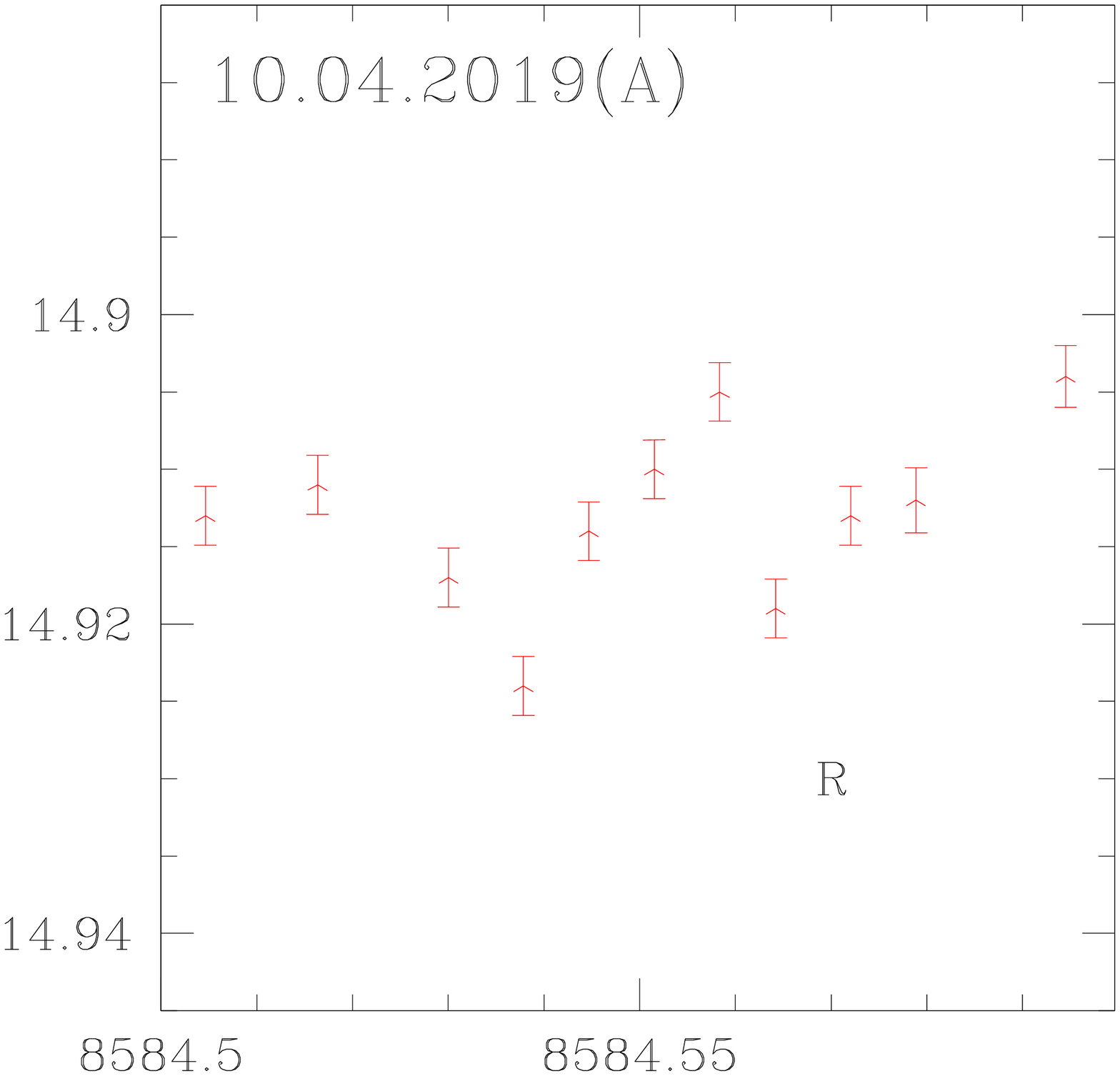,height=1.8in,width=2in,angle=0}
%\hspace{0.3in}
\epsfig{figure=  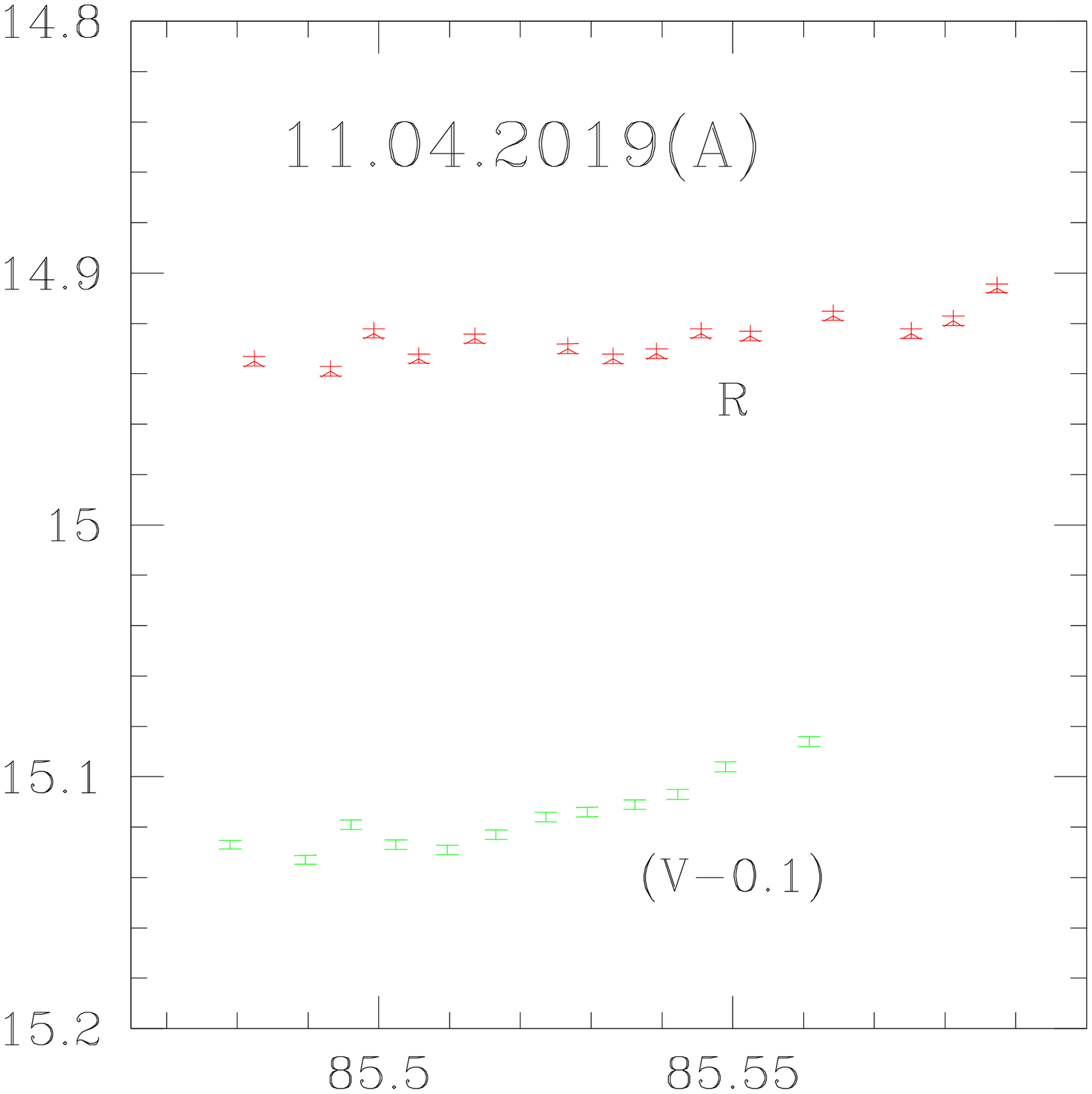,height=1.8in,width=2in,angle=0}
\epsfig{figure=  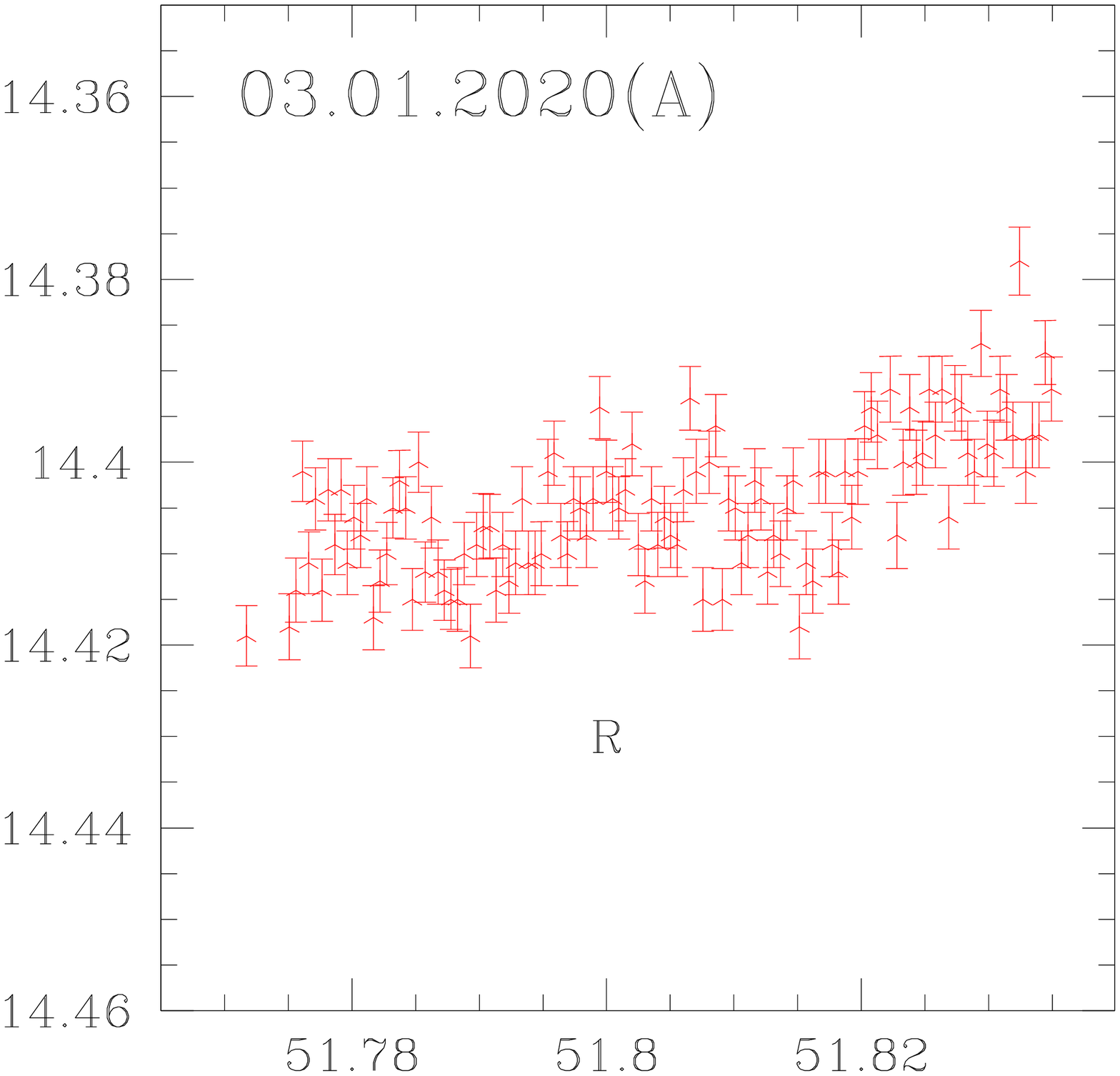,height=1.8in,width=2in,angle=0}

  \caption{Light curves for OJ\,287; green denotes $V$ band while red denotes $R$ filter. In each plot, X axis is JD and Y axis is the source magnitude.
Observation date and the telescope used are indicated in each plot viz.
Telescope A, is JS (2.15\,m Jorge Sahade telescope),  telescope B, HSH (Helen Sawyer Hogg telescope), and
telescope C, JCBT (Vainu Bappu Observatory (VBO), India).}
\label{LC_BL1}
\end{center}
\end{figure*}

\section{Results}
A detailed temporal and spectral study has been done using the multi-wavelength data from the \textit{Fermi}-LAT, Swift-XRT/UVOT telescope. The archival data from OVRO are used to perform the correlation study with the gamma-ray.

\subsection{Multi-waveband Fractional and Temporal Variability}
\subsubsection{Multi-waveband Variability}
OJ 287 is mentioned as a gamma-ray source in 3FGL (\citealt{3FGL}) as well as in 4FGL (\citealt{4FGL}) catalog by Fermi-LAT. The blazar OJ 287 is one of the most active blazars with a binary black hole system, and that makes it one of a kind and thus an interesting source in the Fermi-LAT catalog. It is monitored by various ground-based and space-based telescopes across the entire wavelength range. The recent flare seen by Swift-XRT and UVOT during the beginning of the year 2020 has been confirmed as the second brightest flare in X-ray and optical/UV \citep{Komossa_2020}. 

The multi-wavelength light curve, since January 2017 to May 2020, from radio (at 15 GHz) to gamma-ray (0.1--300 GeV) is shown in Figure \ref{fig:total_lc}.  The whole light curve is divided into various activity states based on the variability and flux states seen in X-ray, optical, and UV. The states are defined as A, B, C, D, and E. Among all these five states, state A and E have higher mag/flux values in optical/X-ray and considered to be flaring states. 
The X-ray flare in state A was earlier studied by
\citet{2018MNRAS.479.1672K}, and \citet{Kapanadze_2018}. They found strong positive correlation between optical, UV and X-ray outbursts. 
Our results are consistent with them, which suggests that the same population of electrons is generating the optical, UV and X-ray outbursts.
The flare in optical and X-ray band during the state E is widely reported in many astronomer's telegram (\citealt{ATel13637}, \citealt{ATel13658}, \citealt{ATel13677}, \citealt{ATel13755}, \citealt{ATel13785}) and studied by \citet{Komossa_2020} and \citet{Kushwaha_2020}. 
%However, the state A is not recognised as higher flux/flaring state earlier and has not been studied in detail. 
Here, we provide a broadband temporal and spectral analysis of these states and the broadband SED modeling is also done to compare the jet parameters between various high (A \& E) and low (B, C, \& D) states.

The upper panel shows the gamma-ray light curve by Fermi-LAT. It is found that the source is not very bright in gamma-ray. The variation in flux is nearly a factor of $5$ between its lowest and highest flux state. The average flux during this period is 2.65$\times$10$^{-8}$ ph cm$^{-2}$ s$^{-1}$. The average flux from the 4FGL catalog for 1$-$100 GeV is 0.6$\times$10$^{-8}$ ph cm$^{-2}$ s$^{-1}$ shown by a horizontal dashed line in Figure \ref{fig:total_lc}. In the last segment of the light curve for the period 2017 - 2020, the gamma-ray data show their highest flux of around 1.0$\times$10$^{-7}$ ph cm$^{-2}$ s$^{-1}$. %The fractional variability and the time variability estimated during all the various states are shown in Table \ref{tab:var}.

 The X-ray light curve is shown in the second panel. It is observed that the source is most variable during the first and fifth states, and the flux became maximum in early 2017. %This appears to be the second-highest flux level reached by OJ 287 in its history \citep{Komossa_2020}. 
The highest flux state in X-ray coincides with the high flux state in radio and optical/UV. In gamma-ray, the flux is not much high, but the variability can be seen in the light curve. 

The light curves for various bands of optical and UV are shown in panels 3rd and 4th. The source seems to be variable across the whole light curve and achieving its maxima in {\bf early 2017 and mid 2020}. These light curves are similar to the X-ray light curve, which suggests the link between their production site and their physical processes. The multi-wavelength SED modeling is done later in this paper to discuss these possibilities. 

The 5th panel shows the simultaneous observations in radio at 15 GHz. The light curve reveals that the source is variable in radio and the maximum variation is a strong decrease from 10 to 1 Jy. \\
The fractional variability estimated for various states is reported in Table \ref{tab:var}.
The fractional variability is used to characterize the long-term variability in various bands. It is formulated by \citet{Vaughan_2003} as:

\begin{equation}
 F_{\rm var} = \sqrt{\frac{S^2 - err^2}{F^2}},
\end{equation}
where, F denotes the mean flux, S$^2$ and err$^2$ are the variance and the mean square error in the flux, respectively. The error in flux variability amplitude is given in \citet{Prince_2019}. The fractional variability amplitude estimated for the various states in all wavebands is depicted in Table \ref{tab:var}. 

The flux doubling/halving time is also estimated for all the states in all bands. The values are further used to characterize the variability in OJ 287. 
The flux doubling time is also known as variability time, is defined as \citep{Zhang_1999}
\begin{equation}
 t_d = \frac{(F_1 + F_2)(T_2 - T_1)}{2|F_2 - F_1|}
\end{equation}
where, F$_1$, and F$_2$ are fluxes at time T$_1$ and T$_2$. The doubling time or the fastest (or shortest) variability time (t$_{\rm var}$) considered as the smallest value among the available pairs in the light curve. 
The variability amplitude and the doubling time together characterize the variability of the source in various states. In segment E, the source appeared to be more variable with the highest variability amplitude among all the states in all wavebands and have the shortest (fastest) variability time of the order of 1 day (Table \ref{tab:var}). The values of variability amplitude vary from 50$\%$ to 60$\%$ among all the wavebands and the shortest variability time of OJ 287 during 2017 to 2020 in X-ray is nearly 1 day.

The radio data are very sparse during this entire period, and hence we did not include them in the variability study to draw any meaningful conclusion.
 
 \subsubsection{Intra-day Variability (IDV)}
 Considering the modest number of observations in each passband, the variability of the source is
measured using the C-criterion, which compares the dispersion in the (blazar-comparison star) and control star -
comparison star. We also used the F test, which is the ratio of the variances of the blazar instrumental light curve (LC) to that of the standard star.
The above tests are discussed in more detail in \citet{2019MNRAS.488.4093A}.
As claimed by
\citet{2017MNRAS.467..340Z}, dispersion scaling by the Howell ($\Gamma_{\rm SF}$) factor \citep{1988AJ.....95..247H} to match the control
star and the target error distributions result in the most reliable results.
We call a particular LC to be variable (Var) only when both tests reject the null hypothesis at 99.5\%
confidence level, possibly variable (PV) if just one of the tests rejects the null hypothesis, and
non-variable (NV) if both tests fail to reject the null hypothesis.
Intraday variability (IDV) results for our observation campaign are summarized in Table\,\ref{tab:var_res} where columns 1 -- 8 are,
respectively, observation date, the passband of observation, number of data points in the given passband, dispersion of
blazar differential LC (DLC), dispersion of the control star DLC, Howell's factor, results for $C$-test and $F$-test.
The variability state of the source is given in column 9.
Our total monitoring coverage contains 13 intraday LCs.
The IDV behavior of OJ 287 over the entire duration is displayed in Figure\,\ref{LC_BL1}.
We found only 1 LC (i.e., April 07, 2019) to be variable according to
our conditions, while 5 LCs were PV. On the remaining seven nights, the source was found to be NV.
Our intraday LCs span a duration of 2 -- 4 hours. Therefore, the relatively short span of observations reduces
the chances of detecting genuine variability. The highest level recorded for OJ 287 was in 2015 Dec by
\citet{2017MNRAS.465.4423G}.
OJ 287 has been monitored for more than a century and has R band data available since 1890. It is one of the
extensively studied sources using both photometry and polarimetry observations on diverse timescales.
During the 2015 flare, the source attained a V mag of $\sim$13.4, R mag of $\sim$ 13.0 while I mag $\sim$ 12.4.
From the current monitoring session, we found the brightest state reached by the source was on January 03, 2020, with R $\sim$ 14.38 mag, fainter than its brightest state in 2015 by $\sim$ 1.4 mag  while the faintest state attained by the source was on December 18, 2019, with
R band mag of 15.15, approximately 2.15 mag fainter than its brightest state during 2015 -- 2016 outburst.
Significant optical LTV is also observed for the source with an R band magnitude change of $\sim$ 2
(Fig.\,\ref{fig:total_lc}). The variability trends observed during our
monitoring period are quite different from the previous ones \citep{2017MNRAS.465.4423G}. The target
did not display high IDV during the current phase, which could be due to the lesser data cadence.
 
\subsection{Gamma-ray Spectral Analysis}
\begin{figure*}
\begin{center}
 \includegraphics[scale=0.35]{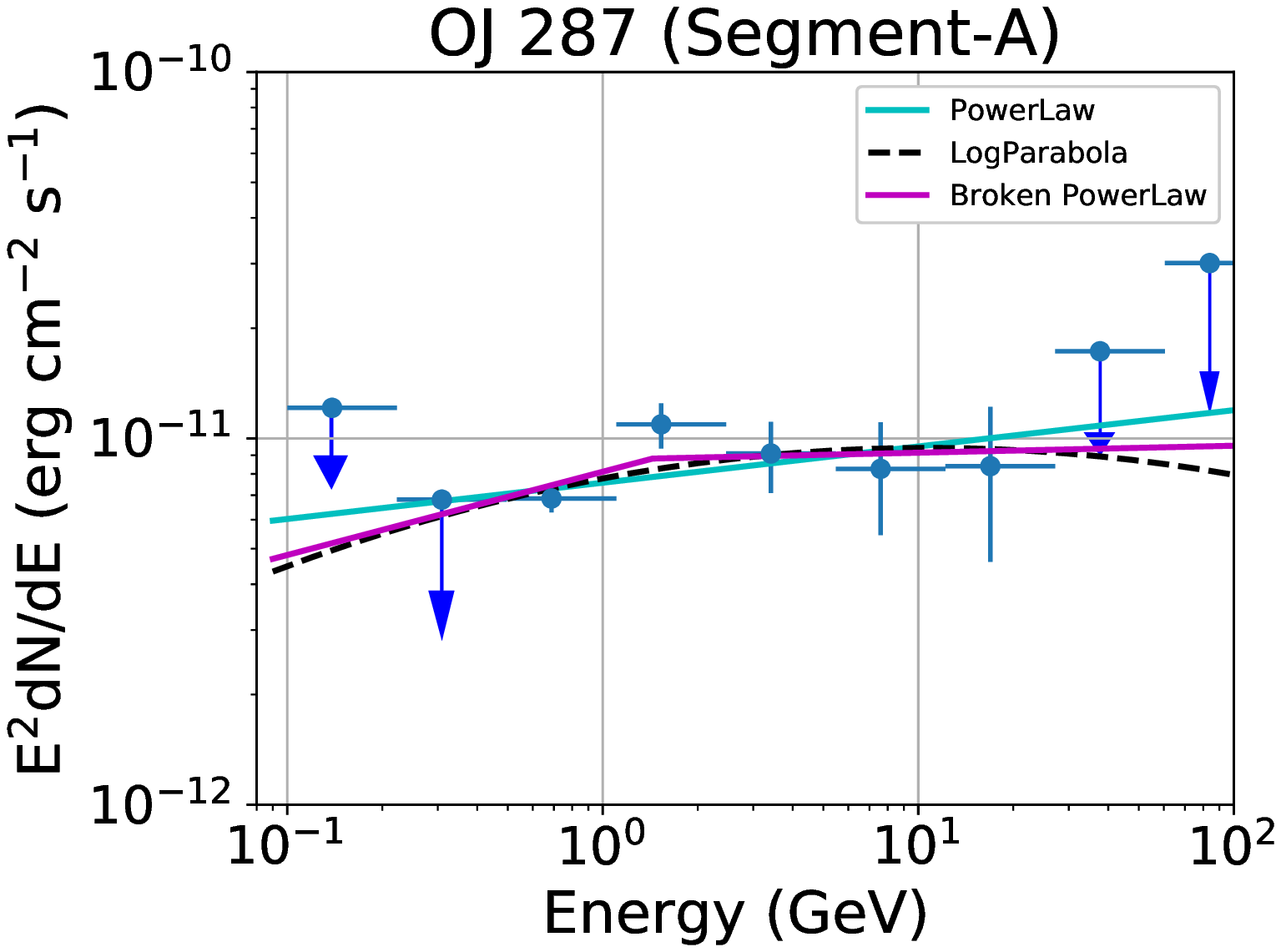}
 \includegraphics[scale=0.35]{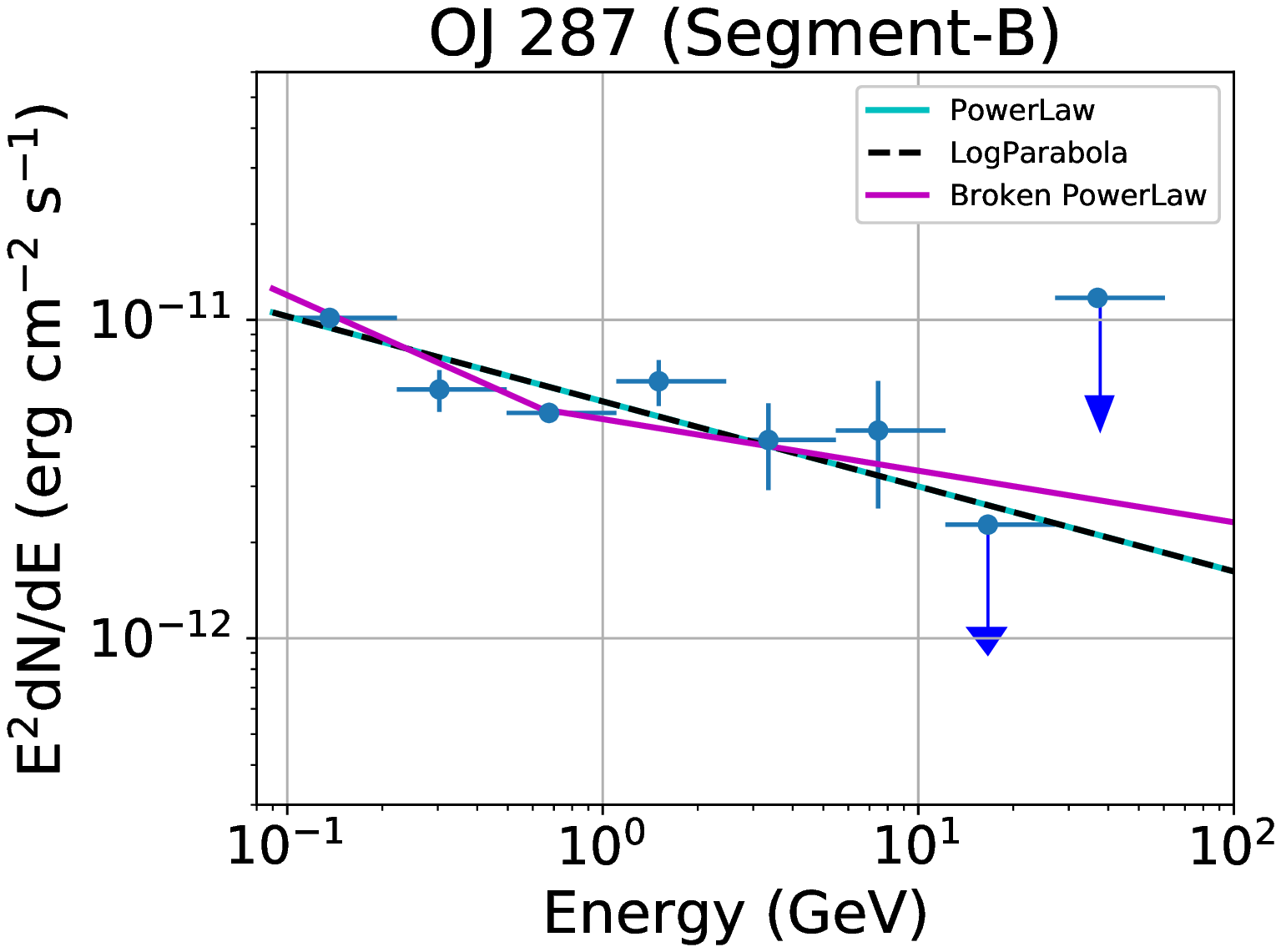}
 \includegraphics[scale=0.35]{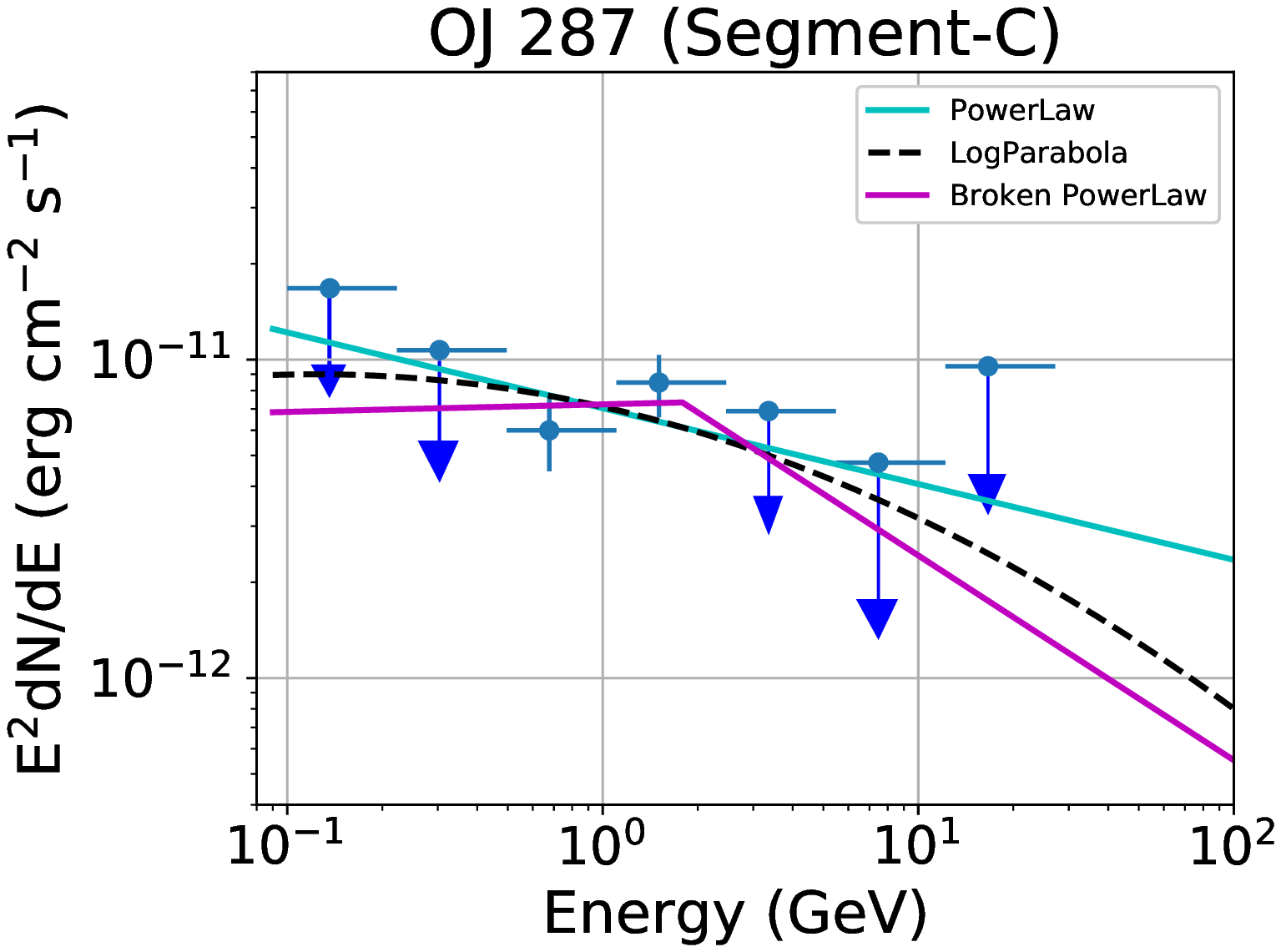}
 \includegraphics[scale=0.35]{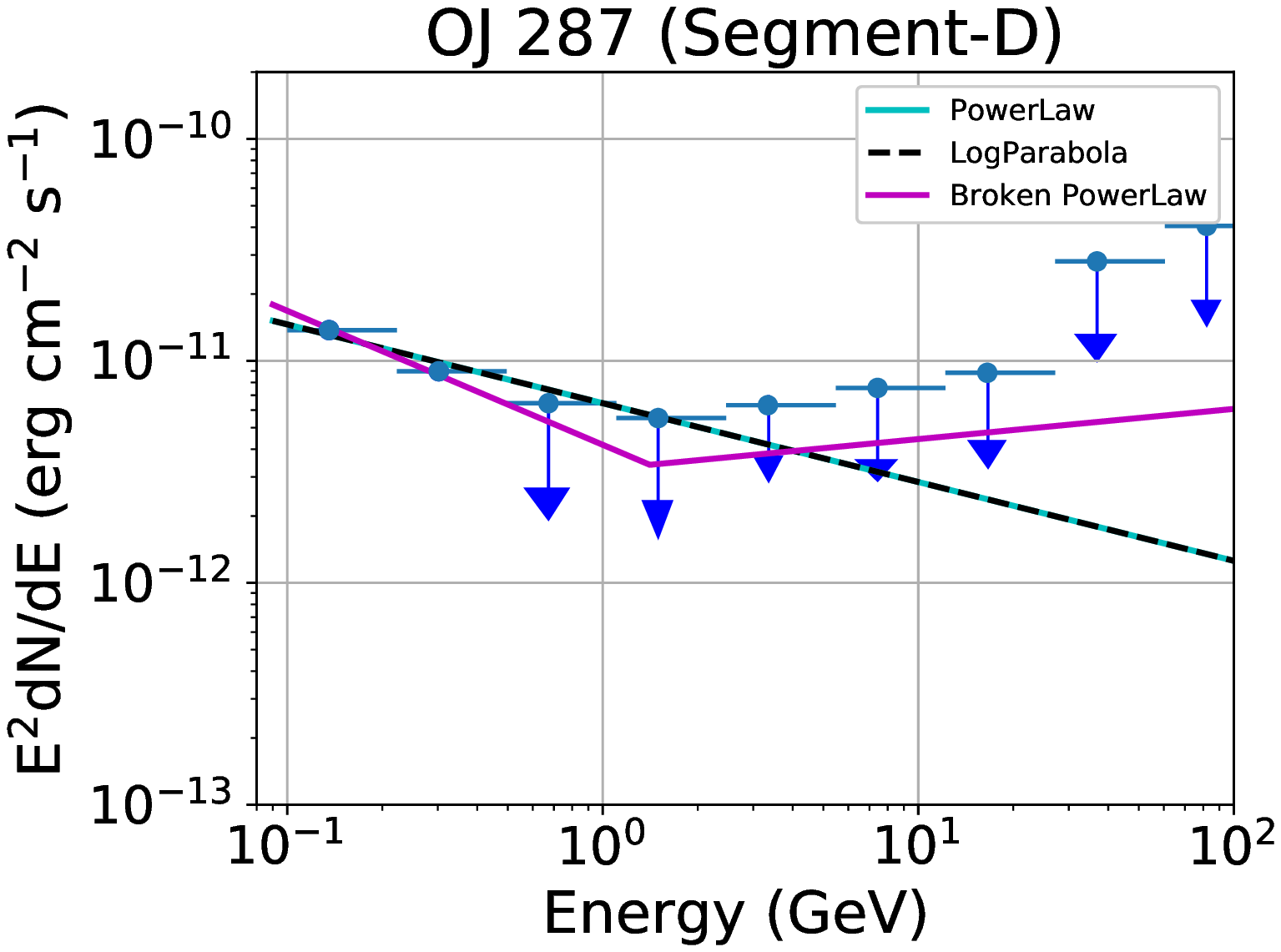}
 \includegraphics[scale=0.35]{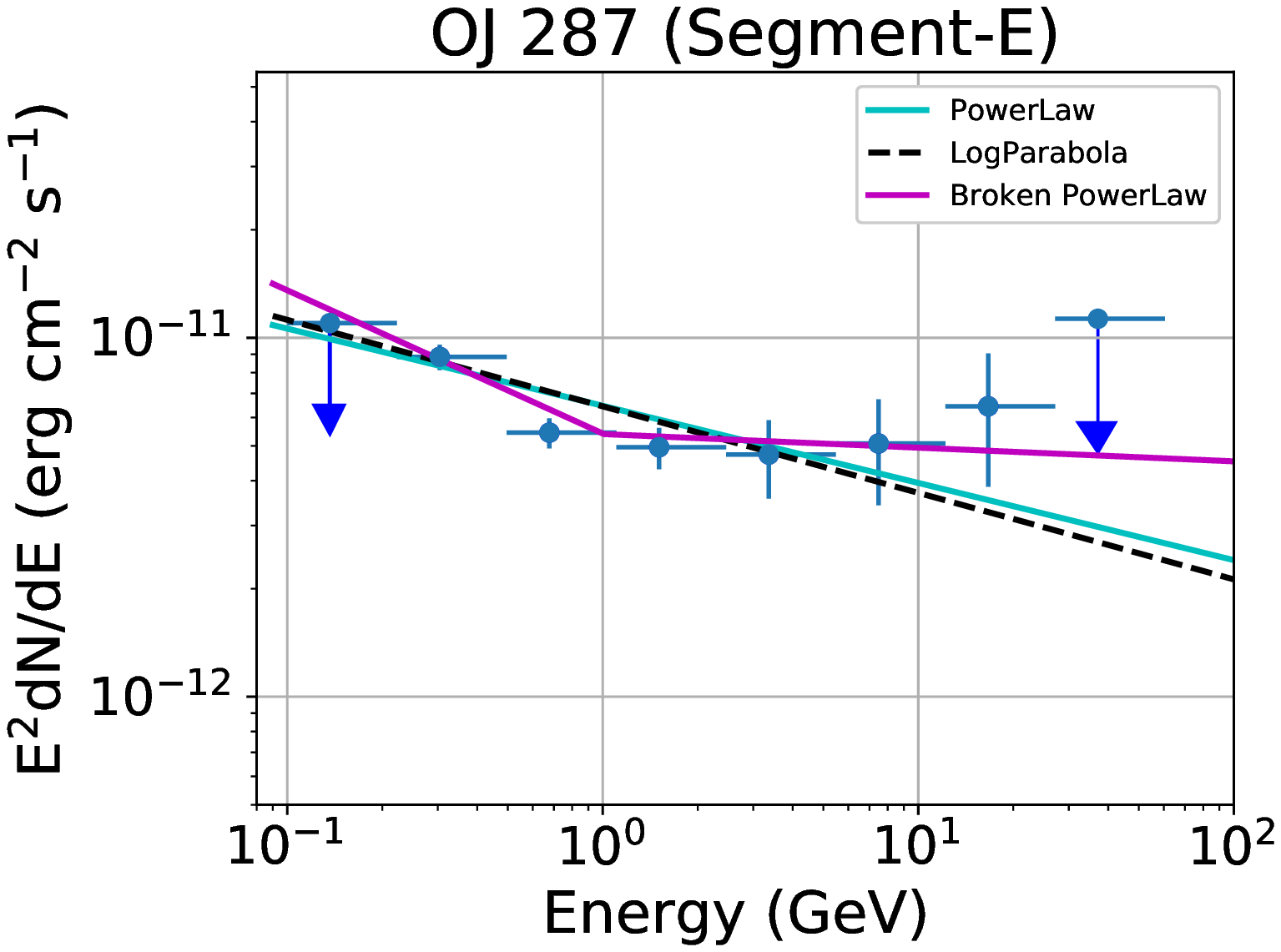}
 \caption{Gamma-ray SED of all the segments identified during 2017$-$2020 in OJ 287 are modeled with three different spectral models PL, LP, and BPL (see Section 3.2 for more details). The down arrow represents the upper limit in that particular segment.}
 \label{fig:gamma_sed1}
 \end{center}
\end{figure*}

We have also produced the gamma-ray spectra for all the various states of the source identified in Figure \ref{fig:total_lc}. The gamma-ray spectra are produced with the help of \textit{likeSED.py}\footnote{https://fermi.gsfc.nasa.gov/ssc/data/analysis/user/} a python code provided by the Fermi Science Tools.
First, the \textit{Likelihood} analysis is done with the default spectral model power-law (PL) to model the spectral data points, and further, we have changed the model to various other models, like log parabola (LP) and broken power law (BPL) to get the best fit. The details about the models are discussed in \citet{Prince_2018}. The isotropic $\gamma$-ray luminosity corresponding to each spectral models are estimated during all the segments by following the equation 5 of \citet{Prince_2021}, and the values are of the order of 10$^{47}$ $-$ 10$^{48}$ erg/s, which are lower than the Eddington luminosity (10$^{50}$ erg/s) of this source as estimated in Section 3.4. The estimated $\gamma-ray$ luminosity values are mentioned in Table \ref{tab:gamma_sed}.
The gamma-ray spectrum and model fitting are shown in Figure \ref{fig:gamma_sed1}, and corresponding model parameters are presented in Table \ref{tab:gamma_sed}.
Considering the PL spectral model, the spectral state of the source changes from segment A to Segment B, C, and D from harder ($\Gamma_{\rm PL}$ = 1.90$\pm$0.06) to softer ($\Gamma_{\rm PL}$ = 2.27$\pm$0.08)) and again it becomes harder from segment D ($\Gamma_{\rm PL}$ = 2.35$\pm$0.10) to segment E ($\Gamma_{\rm PL}$ = 2.21$\pm$0.06).

The \textit{Likelihood} analysis returns the TS (test statistics; TS $\sim$ 25, which corresponds to 5$\sigma$ significance; \citealt{Mattox_1996}) corresponding to each model and is generally used to decide which model gives the best to the spectral data points. So finally, we measure the TS$_{curve}$ = 2(log L(LP/BPL) - log L(PL)), where L represents the likelihood function \citep{Nolan_2012}. The TS$_{curve}$ reveals the presence of curvature or a break in the spectrum, and which could be caused by the absorption of high energy photons ($>$ 20 GeV; \citealt{Liu_2006}) by the broad-line region (BLR), assuming the emitting region is located within the BLR. However, if the emitting region is located outside the BLR, a nice power-law spectral behavior is expected. The best spectral model favors the large positive value of TS$_{curve}$ over the low value of TS$_{curve}$. 

The various models and their corresponding parameters are exhibited in Table \ref{tab:gamma_sed}. The values of TS$_{curve}$  %it would not be fair to comment which model is best describing the data because values are very close to each other, though there is a tuff competition between PL and LP models
are close to each other for the different models. TS values are nearly equal for the PL and LP models but differ with BPL. However, there is no clear trend in the spectral model that can explain the $\gamma$-ray sed from all the segments, which shows that the source has a very complex behavior during 2017$-$2020.  

\begin{table*}
 \centering
 \caption{The modeled parameters of gamma-ray SED for all the segments identified in Figure 1. Column 3 shows the isotropic $\gamma$-ray luminosity during the various segments, which is lower than the Eddington luminosity ($\sim$10$^{50}$ erg/s) of the source as discussed in Section 3.4. }
 \begin{tabular}{c c c c c c c c}
 \hline 
 \noalign{\smallskip}
 Various & F$_{0.1-300 \rm{GeV}}$& Luminosity & PowerLaw & & & TS & TS$_{curve}$ \\
 states & (10$^{-8}$ ph cm$^{-2}$ s$^{-1}$) & (10$^{48}$ erg s$^{-1}$) & $\Gamma_{\rm PL}$ & & & & \\
  \noalign{\smallskip} \hline \noalign{\smallskip}
 A       & 4.10$\pm$0.50&0.25 & -1.90$\pm$0.06 & -- & -- & 570.20 & --\\
 B       & 4.90$\pm$0.60&0.80 & -2.27$\pm$0.08 & -- & -- & 313.00  & --\\
 C       & 5.10$\pm$0.81&0.93 & -2.24$\pm$0.10 & -- & -- & 173.22 & --\\
 D       & 5.40$\pm$0.81&1.04 & -2.35$\pm$0.10 & -- & -- & 178.32 & --\\
 E       & 5.20$\pm$0.55&2.92 & -2.21$\pm$0.06 & -- & -- & 551.28 & -- \\
\noalign{\smallskip} \hline \noalign{\smallskip}
    && & LogParabola & \\
    & && $\alpha$ & $\beta$ &  & & \\
 \noalign{\smallskip} \hline \noalign{\smallskip} 
 A       & 3.60$\pm$0.70&0.22 & 1.81$\pm$0.12 & 0.03$\pm$0.03 & -- & 569.48 & -0.72 \\
 B       & 4.90$\pm$0.60&0.81 & 2.27$\pm$0.08 & 0.00$\pm$0.00 & -- & 313.01  & 0.01 \\
 C       & 4.70$\pm$0.96&0.84 & 2.19$\pm$0.13 & 0.05$\pm$0.07 & -- & 173.64 & 0.42 \\
 D       & 5.40$\pm$0.81&0.96 & 2.35$\pm$0.10 & 0.00$\pm$0.00 & -- & 178.30 & 0.02 \\
 E       & 5.20$\pm$0.55&2.45 & 2.21$\pm$0.06 & 0.00$\pm$0.00 & -- & 551.24& -0.08 \\  
\noalign{\smallskip} \hline \noalign{\smallskip}
   && & Broken PowerLaw & & E$_{break}$& &  \\
   & && $\Gamma_1$ & $\Gamma_2$ & [GeV] & & \\
\noalign{\smallskip} \hline  \noalign{\smallskip}
 A & 3.70$\pm$0.70&0.21 & -1.77$\pm$0.17 & -1.98$\pm$0.11 & 1.43$\pm$0.74 & 569.44 & -0.76 \\
 B & 5.30$\pm$0.30&1.03 & -2.44$\pm$0.83 & -2.16$\pm$0.09 & 0.64$\pm$0.25 & 315.26  & 2.26 \\
 C & 3.60$\pm$1.20&1.08 & -1.97$\pm$0.24 & -2.64$\pm$0.36 & 1.78$\pm$0.16 & 144.38 & -28.84 \\
 D & 6.20$\pm$0.87&1.16 & -2.64$\pm$0.14 & -1.86$\pm$0.16 & 1.41$\pm$0.13 & 186.32 & 8.00 \\
 E & 5.80$\pm$2.00&3.02 & -2.40$\pm$0.45 & -2.04$\pm$0.06 & 1.00$\pm$0.32 & 560.03&  8.75\\
\noalign{\smallskip} \hline   \noalign{\smallskip}
 \end{tabular}
 \label{tab:gamma_sed}
\end{table*}

\subsection{Correlations Studies}
We have collected multi-wavelength data from 2017 to May 2020. Five different states are identified based on the flux and variability seen in X-rays and Optical/UV.  During this period between January 2017 to May 2020, we have not observed any flare in gamma-ray, but the source is variable in this low state, as can be seen in the top panel of Figure \ref{fig:total_lc}. The source is flaring in X-ray and optical/UV and appears to be more variable in X-rays, optical, UV, and radio (15 GHz) wavebands, as shown in Figure \ref{fig:total_lc} from top to bottom.
Here we try to investigate the correlation between the X-ray and optical/UV emission for all the states since they have good coverage in all the wavebands. The observed time lags between light curves at different wavebands can be really helpful to locate their emission regions along the jet axis. 

\begin{figure*}
\centering
 \includegraphics[scale=0.55]{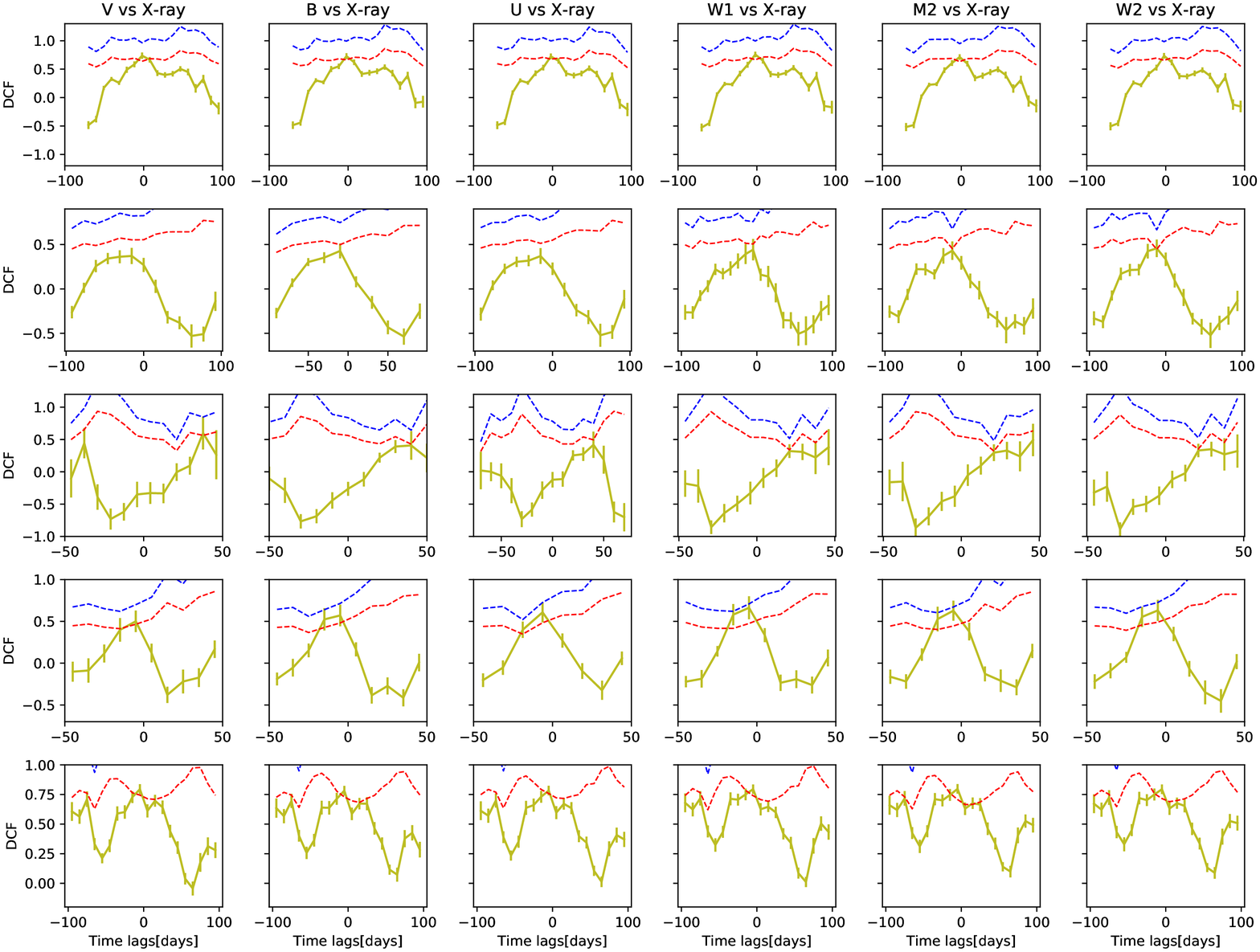} 
 \caption{Cross-correlation of optical/UV vs X-ray. The five rows of the plot are the five states (A, B, C, D, E) defined in Figure \ref{fig:total_lc}. Horizontal red and blue dashed lines are 2$\sigma$ and 3$\sigma$ significance respectively. We do not observe any significant time lag in any of the combinations. During state C, we see some time lags but the correlation coefficient are below 50$\%$ and below 2$\sigma$ significance, and hence do not consider as an actual time lag. }
 \label{fig:dcf}
\end{figure*}

To estimate the correlation, we have followed the method developed by \citet{Edelson_1988}. Different bin sizes have been chosen in different combinations to examine the discrete correlation function (DCF) peaks. The DCF estimated for all the possible combinations between X-rays and UV/Optical are shown in Figure \ref{fig:dcf}. The top row of the figure shows the DCF for state A and then followed by B, C, D, and E towards the bottom row. The correlation coefficients, time lags, and the bin size for all the combinations are mentioned in Table \ref{tab:dcf_tab}. Our results show that optical-X-ray and the UV-X-ray emissions for the states A, B, D, and E are highly correlated with values of correlation coefficient above 50$\%$ and with time lags within the bin size. This strong correlation with zero time lag suggests that these two emissions have a common emission region. However, for state C, we do not observe any correlation between Optical/UV and X-rays. 
We have also estimated the significance of the DCF peaks by simulating the 1000 artificial optical and UV light curves by following the monte carlo procedure described in \citet{10.1093/mnras/stt764}, for PSD slope 1.5. The simulated light curves are cross-correlated with the observed X-ray light curves. Further, 2$\sigma$ and 3$\sigma$ significance is estimated, which are shown in red and blue dashed line in Figure \ref{fig:dcf}. Our results show that in most of the cases, emissions are correlated above 2 $\sigma$ significance.   

As can be seen in Figure \ref{fig:total_lc}, the radio data are very sparse, and hence we did not include it in the correlation study.

\begin{table}
 \centering
 \caption{DCF parameters for all the combinations. Most of the time lags were found within the binsize.}
 \begin{tabular}{c c c c c}
 \hline \noalign{\smallskip}
 States & Combinations & DCF  & Time lags & binsize \\
 \noalign{\smallskip} \hline \noalign{\smallskip}
 A      & V vs X-rays  &0.73$\pm$0.05& -2.06& 10.0  \\
        & B vs X-rays  &0.73$\pm$0.05& -2.06& 10.0 \\
        & U vs X-rays  &0.72$\pm$0.05& -2.06& 10.0 \\
        & W1 vs X-rays &0.76$\pm$0.05& -2.06& 10.0  \\
        & M2 vs X-rays &0.71$\pm$0.05& -2.06& 10.0  \\
        & W2 vs X-rays &0.74$\pm$0.05& -2.06& 10.0  \\
 \noalign{\smallskip}       \hline \noalign{\smallskip}
 B  & V vs X-rays  & 0.37$\pm$0.09&-15.42 & 15.0  \\
    & B vs X-rays  &0.43$\pm$0.08&-10.00 & 20.0  \\
    & U vs X-rays  &0.37$\pm$0.09&-15.42 & 15.0  \\ 
    & W1 vs X-rays &0.44$\pm$0.12&-5.00& 10.0  \\
    & M2 vs X-rays &0.43$\pm$0.10&-11.75& 12.0  \\
    & W2 vs X-rays &0.46$\pm$0.10&-11.75& 12.0  \\
  \noalign{\smallskip}  \hline  \noalign{\smallskip}
 C      & V vs X-rays  &0.60$\pm$0.25&37.60& 8.0  \\
        & B vs X-rays  &0.41$\pm$0.22&40.00& 10.0  \\
        & U vs X-rays  &0.42$\pm$0.20&40.00& 10.0  \\
        & W1 vs X-rays &0.38$\pm$0.28&46.00& 8.0  \\
        & M2 vs X-rays &0.49$\pm$0.25&46.00& 8.0  \\
        & W2 vs X-rays &0.35$\pm$0.11&29.27& 8.0  \\
  \noalign{\smallskip}      \hline  \noalign{\smallskip}
 D      & V vs X-rays  &0.50$\pm$0.13&-5.00&10.0  \\         & B vs X-rays  &0.57$\pm$0.13&-5.00& 10.0  \\
        & U vs X-rays  &0.61$\pm$0.12&-6.28& 12.0  \\
        & W1 vs X-rays &0.66$\pm$0.14&-5.00& 10.0  \\
        & M2 vs X-rays &0.63$\pm$0.13&-5.00& 10.0  \\
        & W2 vs X-rays &0.63$\pm$0.12&-5.00& 10.0  \\
  \noalign{\smallskip}      \hline    \noalign{\smallskip}
  E     & V vs X-rays  &0.79$\pm$0.04&-5.00& 10.0  \\
        & B vs X-rays  &0.78$\pm$0.04&-5.00& 10.0  \\
        & U vs X-rays  &0.78$\pm$0.04&-5.00& 10.0  \\
        & W1 vs X-rays &0.79$\pm$0.04&-5.00& 10.0  \\
        & M2 vs X-rays &0.80$\pm$0.04&-5.00& 10.0  \\
        & W2 vs X-rays &0.79$\pm$0.04&-5.00& 10.0  \\
 \noalign{\smallskip}       \hline   \noalign{\smallskip}
 \end{tabular}
 \label{tab:dcf_tab}

\end{table}

\subsection{Modeling the Multi-wavelength SEDs}
Good coverage of OJ 287 in various wavebands provides an opportunity to obtain the multi-wavelength spectral energy distribution (MWSED), which has been used in our modeling. We have produced the MWSED using {\it Swift}-XRT, UVOT and {\it Fermi}-LAT data for all the observations in different states. The modeling of OJ 287 has been done previously in various ways. 
We assume that the emission region is in the jet of the primary black hole. The emission region is a spherical blob that is moving with Doppler factor $\delta$ down the jet. The shock accelerated leptons are losing energy inside this blob by synchrotron, and synchrotron self Compton (SSC) processes.

We have used a publicly available time-dependent code, GAMERA\footnote{http://joachimhahn.github.io/GAMERA} (\citealt{Hahn_2015}) to model the broadband SED. It is a python based code and needs an initial injected electron spectrum as an input which further solves the transport equation (3) and estimates the propagated electron spectrum. Finally, the propagated electron spectrum is used to calculate the emission from the various processes like Synchrotron, SSC, and  EC by external photons of various origin (BLR, DT, accretion disk).
We use the following transport equation to find the electron spectrum after energy loss:

\begin{equation}\label{8}
\frac{\partial N(E,t)}{\partial t}=Q(E,t)-\frac{\partial}{\partial E}\Big(b(E,t) N(E,t)\Big)
\end{equation}
where, $Q(E,t)$ is the input spectrum and $N(E,t)$ is the propagated one at a time `t'. 
$b(E,t)$ corresponds to the radiative loss by different physical processes, synchrotron, SSC, and EC scattering. We have assumed a LogParabola electron distribution as the injected electron spectrum in our modeling.

 The MWSED could be modelled with the leptonic scenario 
 where SSC (synchrotron self Compton) and EC (external Compton) emission are generating the high energy peak. The study by \citealt{Kushwaha_2013} on the 2009 flare suggests that the X-ray and $\gamma$-ray emission can be explained by SSC and EC processes respectively, where the seed photons for the external Compton are originated by a thermal bath of 250 K located at distance of $\sim$9 pc from the SMBH.
 In a more recent study by \citealt{2018MNRAS.473.1145K} the December 2015 - May 2016 high state has been modelled using both SSC and EC emissions. 
 The December 2015 high activity has been predicted to occur from the impact of the second black hole on the accretion disk of the primary black hole. The non thermal emission showed nearly co-spatial origin. They modelled the gamma ray flux by external Compton emission of relativistic electrons by optical-UV line emission, which shows the signature of a blue bump in the optical-UV flux. 
 They fitted the X-ray data with SSC emission.
 In the present study we find that the source is variable in all wavebands including $\gamma$-ray, though we did not see any flaring behavior in this band. The observed day scale variability time in gamma-ray flux suggests the 
 location of the emission region close to the SMBH, within a few parsec from the base of the jet. The variability study across all the wavebands exhibits an order of day scale flux variability, due to this reason we model the broadband SED with only synchrotron and SSC within a single emission zone. The correlation study also suggests that the emissions are produced at the same location.

The SSC emission is determined by the synchrotron emission and the size of the emission region. The synchrotron emission depends on the magnetic field in the emission region in the jet and the energy of the leptons.
The size of the emission region can be constrained by the variability time scale. Considering 1 day variability time in the gamma-ray data, we have estimated the size of the emission region by the following  relation r$\sim$ c t$_{var}$ $\delta$/(1+z), where $\delta$ = 20, and it is found to be r$\sim$4.0$\times$10$^{16}$cm.

 However, in our modeling the size of the emission region 
is a free parameter and we have found that a smaller size of the emission region is required to explain the broadband SED. 

\begin{figure*}
 \includegraphics[scale=0.4]{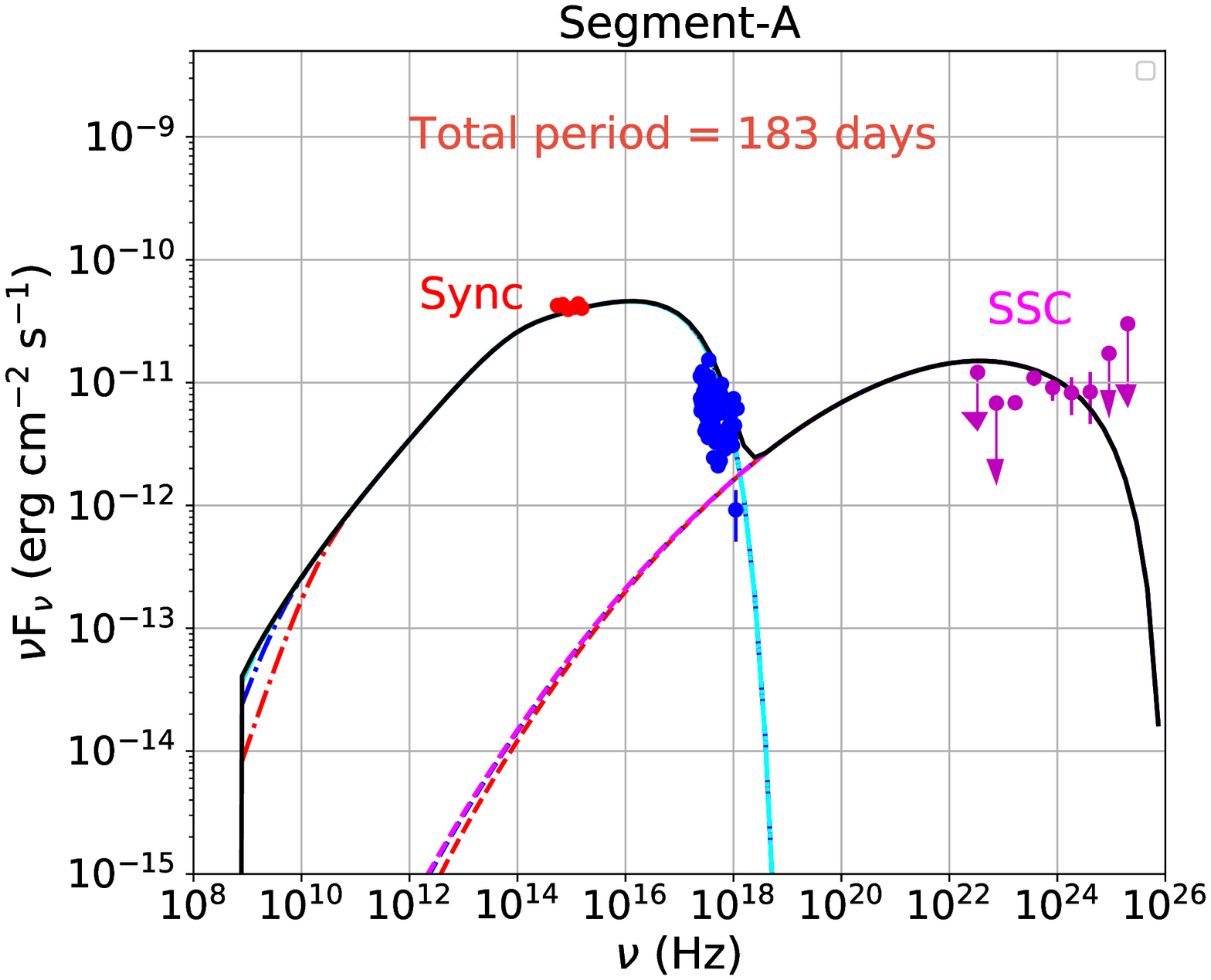}
 \includegraphics[scale=0.4]{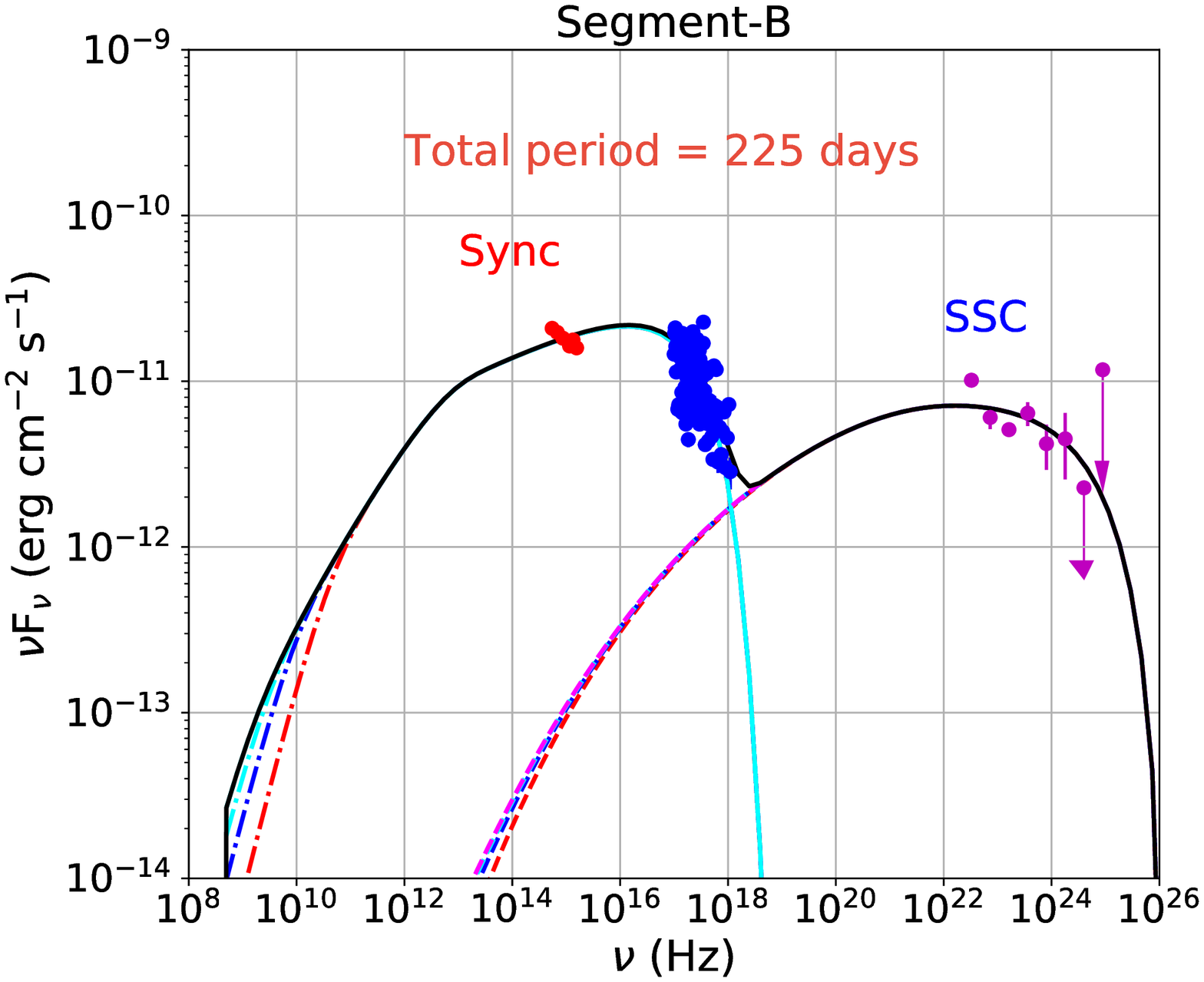}
 \includegraphics[scale=0.4]{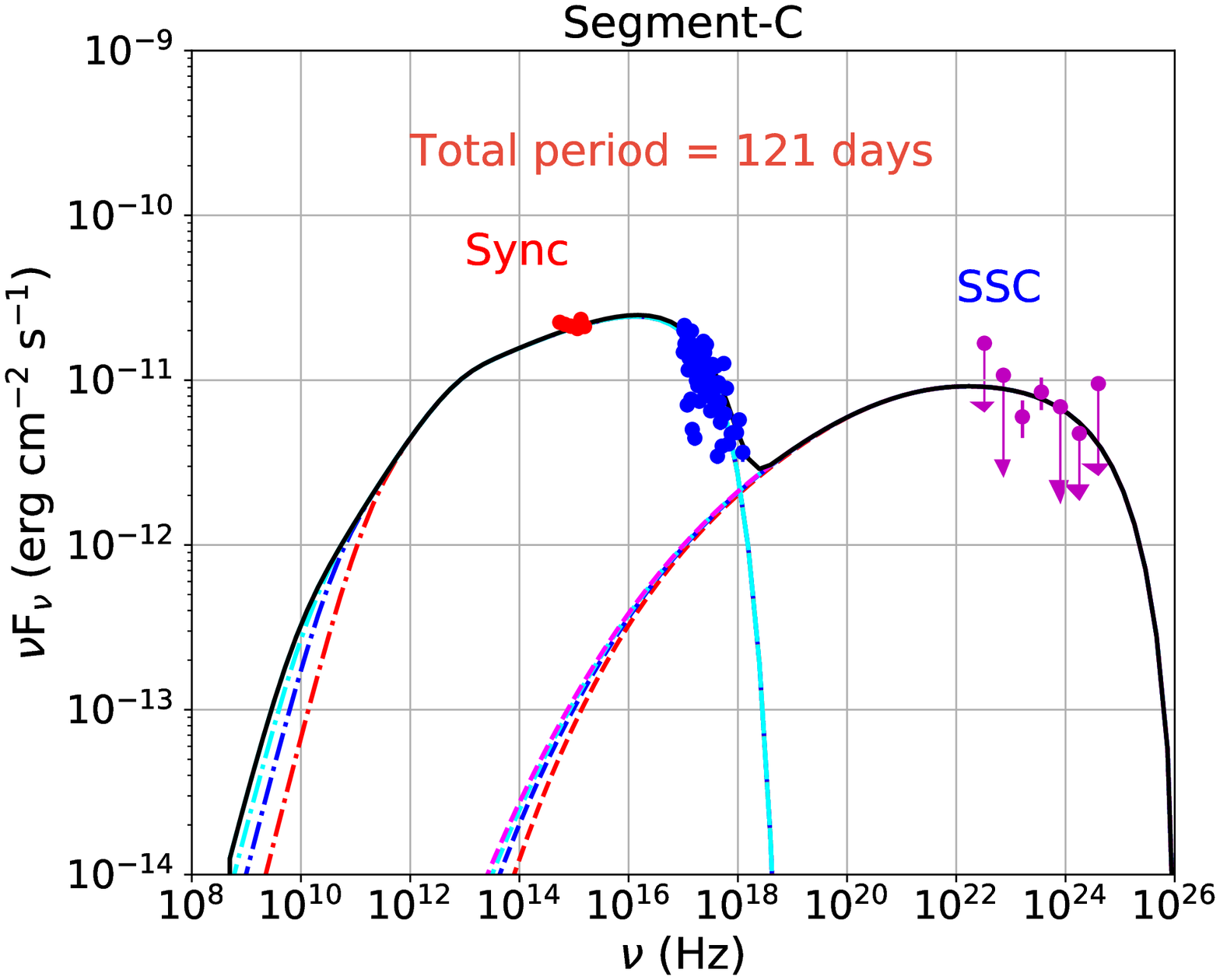}
 \includegraphics[scale=0.4]{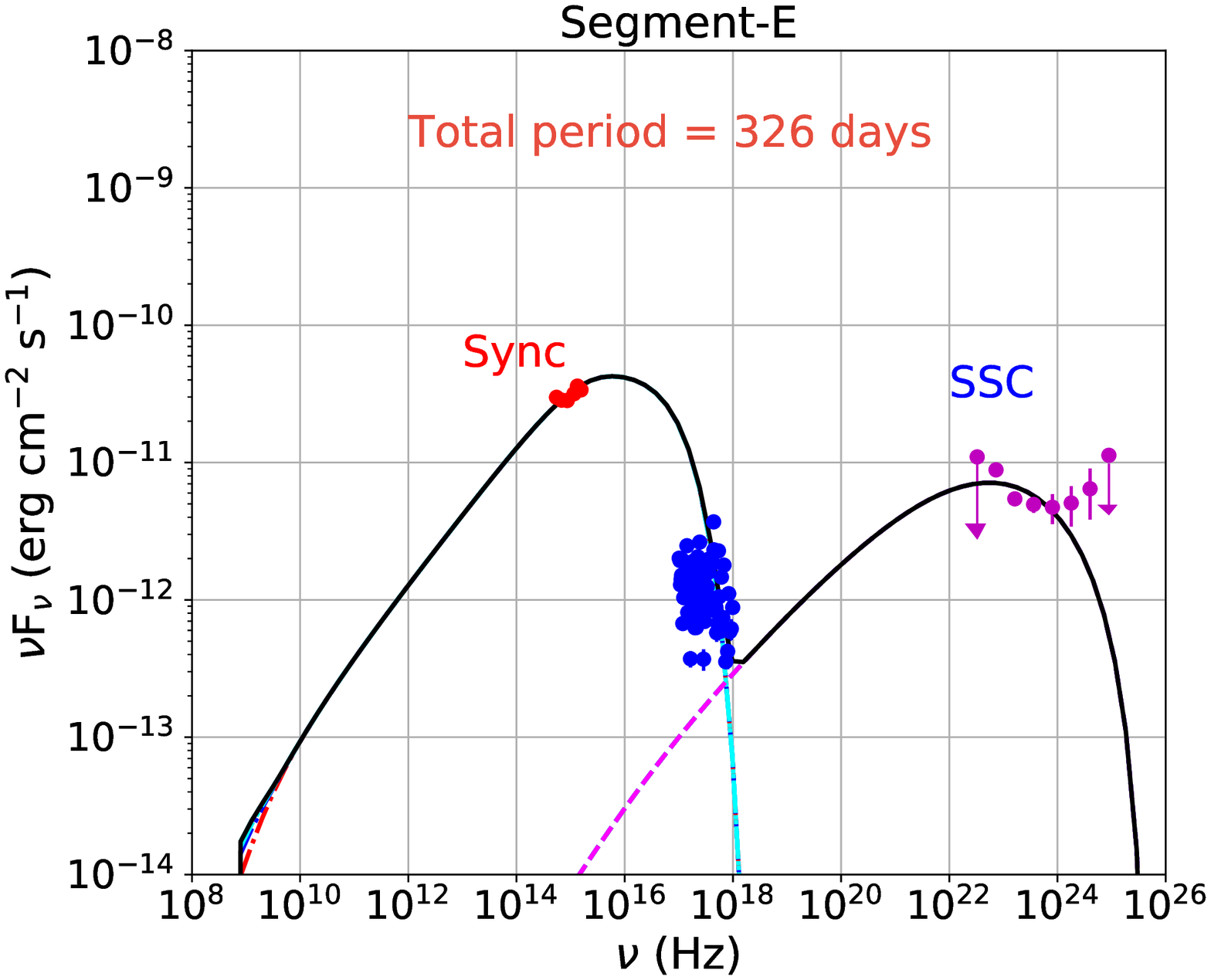}
 \caption{ The MWSED for all the various segments observed during the year 2017$-$2020. The `dotted dahsed' and `dahsed' line in different colors in Synchrotron and SSC peaks are the time evolution of the model. The down arrow represents the $\gamma$-ray upper limits. The optical/UV, X-ray, and $\gamma$-ray data points are shown in red, blue, and magenta colors respectively.}
 \label{fig:MWSED}
\end{figure*}

 We also have many other parameters in our model, e.g., magnetic field, injected electron spectrum, lower and higher energy cut-offs in the injected electron spectrum, normalization of the electron spectrum, and these are optimized to achieve the best SED fit.
The MWSED modeling results are depicted in Figure \ref{fig:MWSED} for the various states, and their corresponding best-fit parameters are tabulated in Table \ref{tab:sed_param}. States C $\&$ D are very similar to each other in variability and flux states, and hence we only show the SED modeling of state C.
The modeling confirms that the low and high energy peaks can be constrained with synchrotron and SSC, respectively.
%EC processes by thermal bremsstrahlung photons, respectively. 
In Figure \ref{fig:MWSED}, it has been noticed that the source has more emission in Optical-UV than in $\gamma$-ray. Hence a large value of magnetic field ($\sim$4-7 Gauss) is used to fit the data. 

 The previous broadband SED modeling of OJ 287 at different occasions of low and bright state  (\citealt{Kushwaha_2013}, \citealt{2018MNRAS.473.1145K, 2018MNRAS.479.1672K}) was carried out with a 
range of values of the Doppler and Lorentz factor.  In this study, we have fixed the Doppler and Lorentz factor of the blob at 20 and 15.5 respectively, which are similar to the values reported in earlier papers.

\begin{table*}
\centering
\caption{Multi-wavelength SED modeling results with the best fitted parameters values. The input injected electron distribution is LogParabola with reference energy 60 MeV. The Doppler factor and the Lorentz factor are fixed at 20.0 and 15.5 respectively. }
 \begin{tabular}{c c c c c}
 \hline \noalign{\smallskip}
 high state& Parameters & Symbols & Values & Period \\
\noalign{\smallskip}  \hline  \noalign{\smallskip}
 Segment-A & &&& 183 days\\
 & Size of the emitting zone& r & 2.6$\times$10$^{15}$ cm & \\
 & Min Lorentz factor of emitting electrons & $\gamma_{min}$& 350.0 &\\
 & Max Lorentz factor of emitting electrons & $\gamma_{max}$& 2.8$\times$10$^{4}$ &\\
 & Input injected electron spectrum (LP) & $\alpha$ & 1.60 & \\
 & Curvature parameter of the PL spectrum & $\beta$& 0.02 & \\
 & Magnetic field in emitting zone & B & 5.9 G & \\
 & Jet power in electrons & P$_{j,e}$ & 4.35$\times$10$^{44}$ erg/s & \\
 & Jet power in magnetic field & P$_{j,B}$ & 2.12$\times$10$^{44}$ erg/s & \\
 & Jet power in protons & P$_{j,P}$ & 3.39$\times$10$^{43}$ erg/s& \\
 & Total jet power & P$_{jet}$ & 6.81$\times$10$^{44}$ erg/s& \\
 \noalign{\smallskip} \hline   \noalign{\smallskip}
Segment-B & &&& 225 days \\
& Size of the emitting zone& r & 2.6$\times$10$^{15}$ cm & \\
 & Min Lorentz factor of emitting electrons & $\gamma_{min}$& 120.0 &\\
 & Max Lorentz factor of emitting electrons & $\gamma_{max}$& 3.6$\times$10$^{4}$ &\\
 & Input injected electron spectrum (LP) & $\alpha$ & 1.68 & \\
 & Curvature parameter of the PL spectrum & $\beta$& 0.005 & \\
 & Magnetic field in emitting zone & B & 4.2 G & \\
 & Jet power in electrons & P$_{j,e}$ & 2.59$\times$10$^{44}$ erg/s & \\
 & Jet power in magnetic field & P$_{j,B}$ & 1.07$\times$10$^{44}$ erg/s & \\
 & Jet power in protons & P$_{j,P}$ & 3.56$\times$10$^{43}$ erg/s& \\
 & Total jet power & P$_{jet}$ & 4.02$\times$10$^{44}$ erg/s& \\
 \noalign{\smallskip}  \hline  \noalign{\smallskip}
 Segment-C & &&& 121 days \\
 & Size of the emitting zone& r & 2.6$\times$10$^{15}$ cm & \\
 & Min Lorentz factor of emitting electrons & $\gamma_{min}$& 160.0 &\\
 & Max Lorentz factor of emitting electrons & $\gamma_{max}$& 3.6$\times$10$^{4}$ &\\
 & Input injected electron spectrum (LP) & $\alpha$ & 1.68 & \\
 & Curvature parameter of the PL spectrum & $\beta$& 0.005 & \\
 & Magnetic field in emitting zone & B & 4.2 G & \\
 & Jet power in electrons & P$_{j,e}$ & 2.96$\times$10$^{44}$ erg/s & \\
 & Jet power in magnetic field & P$_{j,B}$ & 1.07$\times$10$^{44}$ erg/s & \\
 & Jet power in protons & P$_{j,P}$ & 4.06$\times$10$^{43}$ erg/s& \\
 & Total jet power & P$_{jet}$ & 4.44$\times$10$^{44}$ erg/s& \\
\noalign{\smallskip} \hline  \noalign{\smallskip}
 Segment-E & &&& 326 days \\
 & Size of the emitting zone& r & 2.6$\times$10$^{15}$ cm & \\
 & Min Lorentz factor of emitting electrons & $\gamma_{min}$& 1.4$\times$10$^{3}$ &\\
 & Max Lorentz factor of emitting electrons & $\gamma_{max}$& 1.5$\times$10$^{4}$ &\\
 & Input injected electron spectrum (LP) & $\alpha$ & 1.6 & \\
 & Curvature parameter of the PL spectrum & $\beta$& 0.005 & \\
 & Magnetic field in emitting zone & B & 6.7 G & \\
 & Jet power in electrons & P$_{j,e}$ & 3.26$\times$10$^{44}$ erg/s & \\
 & Jet power in magnetic field & P$_{j,B}$ & 2.73$\times$10$^{44}$ erg/s & \\
 & Jet power in protons & P$_{j,P}$ & 1.32$\times$10$^{43}$ erg/s& \\
 & Total jet power & P$_{jet}$ & 6.12$\times$10$^{44}$ erg/s& \\
\noalign{\smallskip}  \hline  \noalign{\smallskip}
 \end{tabular}
 \label{tab:sed_param}

\end{table*}

We have also estimated the total jet power and the power in the individual component of the jet. The components are leptons, magnetic fields, and protons. We assume that the number ratio of leptons to protons is 20:1 in the jet and estimate the jet power in leptons and protons separately.  
The total jet power is generally defined as,
\begin{equation}
 P_{jet} = \pi r^2 \Gamma^2 c (U'_e + U'_B + U'_p)
\end{equation}
where, $U'_e$, $U'_B$, and $U'_p$ are the energy densities in leptons, magnetic field and protons in the jet or co-moving frame. The values of the size of the emission region (`$r$') and the Lorentz factor ($\Gamma$) are already provided in the discussion above. 

The total jet power calculated for all the states is shown in Table \ref{tab:sed_param}, and the value is much smaller than the Eddington luminosity of the source. The Eddington luminosity for the primary BH is estimated as L$_{Edd}$ = 4$\pi$Gmm$_p$c/$\sigma_T$, where `m' is the mass of the primary BH, m$_p$ is the proton mass, and $\sigma_T$ is the Thompson scattering cross-section. The primary BH mass is estimated by \citet{2018MNRAS.473.1145K} by modeling the NIR-optical spectrum with an accretion disk, and the reported value is $\sim$1.8$\times$10$^{10}$M$_\odot$. The Eddington luminosity is estimated to be 2.30$\times$10$^{50}$ erg/s, which is much higher than the total jet power estimated in this study by SED modeling.
 Modeling the high optical flux state with synchrotron emission requires a higher value of the magnetic field and hence higher jet power in the magnetic field. It is found that during the flaring state A \& E the total jet power is 1.5 times higher than the total jet power estimated for low state B \& C. The SED modeling also suggests that more luminosity in high energetic electrons is required to produce the broadband emission during flaring states A \& E. 
The non thermal flares during state A and state E might have resulted from disk impact in November-Dec 2015 (\citealt{Valtonen_2016}) and July-September 2019  (\citealt{Laine_2020}) respectively when thermal flares were observed.
The injection of high energetic electrons in jet could be due to the time delayed increase in the accretion rate and jet activity triggered by disk impact of secondary black hole or by tidal disruption events \citet{Sundelius_1997}. 
The variable accretion rate causes internal shock in the jet which accelerates electrons and they lose energy radiatively (\citealt{Valtonen_2006}).
The model of \citet{Sundelius_1997} predicted a major increase in accretion rate in beginning of January 2020. However, the non-thermal flares happened during April-June 2020 nearly 4 months after their predicted time. The physical explanation for this delay requires a better understanding of the disk-jet connection as discussed by \citet{Komossa_2020}.

\section{Discussions and  Conclusions}
During the period between 2017 -- 2020, the blazar OJ 287 did not show any bright flaring states in $\gamma$-ray. However, high flux states were reported across the optical-UV and X-ray wavebands in various Atels notifications, during that period. 
Also, variability in flux was observed in optical, UV, X-ray and gamma-ray frequency. 
Five states have been identified as A, B, C, D, \& E based on the flux and fractional variability seen in optical-UV and X-rays. States A and E appear to be the brightest among all others in Figure \ref{fig:total_lc},  which can also be verified by the total jet power (Table \ref{tab:sed_param}) found from the modeling of these states. The variability time found across the bands ranges from 12 hr to $~$ 20 days across all states. The fastest variability time in X-rays is found to be of the order of 1 day. The optical bands U, B, \& V have the shortest variability time of $\sim$ 14 hr, 30 hr, \& 18 hr, while in UV bands, they are of the order of 1 day, 4 days, and 1 day for W1, M2, \& W2 bands respectively. Though the source was not bright in $\gamma$-rays, we have produced the $\gamma$-ray spectrum for the different states to see if there is any variation in the spectrum. 
The gamma-ray data shows day scale variability and the maximum variation in flux between high and low state is found to be 5 times.
The $\gamma$-ray data can be well fitted with PL or LP model. The values of the test statistics are similar for both the models.
Further, we have estimated the correlations between various wavebands in order to understand whether they have a common emission region. The results show that emission is highly correlated (within the errorbars) between the different bands, which suggests their co-spatial origin. A single-zone emission model is applied to explain the multi-wavelength emission by performing the MWSED modeling.  The SED modeling confirms the presence of high magnetic field in the jet and that the jet emission is powered by relativistic electrons.
\par
 In the binary black hole model the primary black hole is surrounded by an accretion disk. The orbit of the secondary black hole around the primary black hole is such that it intersects the accretion disk of the primary black hole two times.
%perpendiculer to the accretion disk and  intersecting it. 
The major outbursts which occur at approximately 12 years of interval could be due to tidally induced mass flows. One such outburst is expected for every pericenter passage of the secondary (\citealt{Sundelius_1997}). \citet{Pihajoki_2013} theoretically predicted the timings of the precursor flares and compared with the observed flares in the light curve of OJ 287. Based on the model of \citet{Sundelius_1997} a major after-flare is expected in January 2020, but it was observed in May 2020.
 The various physical conditions which affect this time delay are disk/corona properties and geometry, magnetic field geometry, shock formation in the jet. They are not yet well understood (\citealt{Komossa_2020}), which makes it very hard to predict this time delay from first principle. 
 \par
 
The disk impact model predicts thermal bremmsstrahlung radiation as outbursts in optical-UV frequency due to the impact of the secondary black hole on the accretion disk of the primary black hole \citet{Lehto_1996}. This model successfully predicted the impact flares
 in 2007, 2015, 2019 (\citealt{Dey_2021}). The disk impact triggers time delayed increase in accretion rate and jet activity which leads to after-flare effects. The flares in state A \& E in our work can be explained with this model. During states B, C, D there was no flare, hence a lower jet power is needed to model these low states. 

\par
Microvariability studies of blazars are one of the most relevant probes to understand the physical conditions very close to the central supermassive black hole. The exact phenomenon behind IDV in blazars is still under debate.
Flux variations in blazars on intraday timescales almost certainly arise from intrinsic factors that are inherent to the blazar jets such as shocks in the helical jets \citep{2015JApA...36..255C} or blobs of plasma traversing through the Doppler boosted magnetized jet or formation of ultra-relativistic mini jets in the helical jet itself. In the low state of blazars, an alternative source for optical IDV is the accretion disc \citep[AD, e.g.][]{1993A&A...271..216C}. According to the AD-based models, instabilities or hot spots or any other enhanced emission on the AD can yield optical IDV in blazars when the source is in a low state. The presence of confirmed IDV on only 1 night out of 13 nights could be most likely due to a uniform jet emission and any change in the direction of the relativistic shock with respect to our line of sight (LOS) if at all present, is very weak.
LTV in blazars can be attributed to both intrinsic and extrinsic factors. Extrinsic mechanisms involve geometrical effects such as deviation of the emitting region with respect to LOS, thus causing variation in the Doppler factor \citep{2009A&A...504L...9V} which in turn is observed as a variation on a long-term basis. Long-term flux variations in blazar LCs can also be caused by the launching of the new shocks. In general, optical IDV in blazars involve both intrinsic and extrinsic mechanisms \citep{2016ApJ...820...12P} and are usually difficult to disentangle.

\citet{Komossa_2020} studied the large X-ray data sample from 2015 to 2020, and their results pose a few fundamental questions. They observed the strong flare in X-ray, optical, and UV bands. The observation at the peak of the X-ray flare shows a steep power-law spectrum with index 2.8, which is very rare in blazars but consistent with the synchrotron origin of X-ray emission. They concluded that the emission is jet-driven and which is consistent with the binary black hole model. In another study by \citet{Kushwaha_2020}, the spectral change in X-ray emission was noted during 2017 to 2020, and they also suggested that it could be an emission from the jet. Here we have modeled the various low and high states observed during 2017 to 2020 considering the emissions are produced inside the jet.

It was reported earlier by \citet{2018MNRAS.473.1145K} that the source was active during December 2015 $-$ April 2016 in IR to $\gamma$-ray frequency. However, another study by \citet{2018MNRAS.479.1672K} for the period of June 2016 $-$ September 2017 found that the source was very bright in IR to X-ray, but no variability was seen in $\gamma$-ray. A similar kind of behavior is seen during early 2017 and mid 2020 when the source is flaring in optical-UV and X-ray but not very active in  $\gamma$-ray. The different behaviors at different frequencies and at different epochs of time make this source very complex in nature. Many more observational and theoretical studies are required to understand the complex nature of the blazar OJ 287. %The study presented in this work is a small step towards understanding this source. 

\section*{Acknowledgements}
The project was partially supported by the Polish Funding Agency National Science Centre, project 2017/26/A/ST9/00756 (MAESTRO 9), and MNiSW grant DIR/WK/2018/12.
Based on data obtained at Complejo Astronómico El Leoncito, operated under agreement between the Consejo Nacional de Investigaciones Científicas y Técnicas de la República Argentina and the National Universities of La Plata, Córdoba and San Juan.

%%%%%%%%%%%%%%%%%%%%%%%%%%%%%%%%%%%%%%%%%%%%%%%%%%



\bibliographystyle{aa}
\bibliography{reference-list.bib} % if your bibtex file is called example.bib
\end{document}